\documentclass[
aps, prd, 
preprint,
lengthcheck,
superscriptaddress,
nobibnotes,
hyperref
]{revtex4-2}

\usepackage{amsmath, amsfonts, amssymb, bbm, bm}
\usepackage{mathtools}
\usepackage{arydshln}   % dashed line
\usepackage{slashed}
\usepackage{enumerate} % item
\usepackage{graphicx} % Required for inserting images
\usepackage{subcaption} % Required for inserting subfigures
\usepackage[percent]{overpic}
\usepackage{color}
\usepackage{cancel} % 删除线 \cancel{a+b=c} 更适合公式
\usepackage{ulem} % 删除线 \sout{text} 更适合文字
\usepackage{hyperref}

\definecolor{purple_main}{RGB}{126, 30, 156}
\definecolor{violet_main}{RGB}{154, 14, 234}
\definecolor{orange_main}{RGB}{249, 115, 6}
\definecolor{slateblue_main}{RGB}{0, 0, 139}

\def\be{\begin{eqnarray}}
	\def\ee{\end{eqnarray}}

\newcommand{\cmb}[1]{{\color{blue}{ #1 }}}

\begin{document}

\title{Finite density nuclear matter and neutron stars in hard-wall AdS/QCD model}

\author{Jun-Shuai Wang}
    \email{wangjunshuai@ucas.ac.cn}
    \affiliation{International Centre for Theoretical Physics Asia-Pacific (ICTP-AP), 
University of Chinese Academy of Sciences (UCAS), Beijing, 100190, China}
    \affiliation{Taiji Laboratory for Gravitational Wave Universe (Beijing/Hangzhou), 
University of Chinese Academy of Sciences (UCAS), Beijing, 100049, China}
\author{Li-Kang Yang}
 \affiliation{ School of Physics, Nanjing University, Nanjing, 210093, China
 }
\author{Yin-Fang Liu}
    \affiliation{
        School of Frontier Sciences, Nanjing University, Suzhou, 215163, China
    }
    
\author{Yong-Liang Ma}
    \email{ylma@nju.edu.cn}
    \affiliation{
        School of Frontier Sciences, Nanjing University, Suzhou, 215163, China
    }
%    \affiliation{\addrIctpAp}

\date{\today}

\begin{abstract}
We investigate properties of nuclear matter, equation of state (EOS) of 
neutron stars and its mass-radius relation in a hard-wall AdS/QCD model by regarding baryons as solitonic configurations in gauge fields. 
Compared with previous approaches, 
we employ a different homogeneous approximation that takes into account the equations of motion for the pure gauge fields. By choosing appropriate parameters, we realize a chiral phase transition within the baryonic phase, where the chiral condensate decreases with the baryon chemical potential, until it reaches zero---chiral symmetry is restored. In addition, independent of  the existence of chiral phase transition, we also find that the speed of sound converges to the conformal limit at the density relevant to cores of massive stars but the trace of energy-momentum tensor does not vanish which indicates the pseudoconformal structure and intrinsic manifestation of scale symmetry in compact star matter. Through calculations, we obtain an equation of state that is more tightly constrained than previous works, and the resulting mass-radius relation of neutron stars is consistent with current observations.
    
\end{abstract}

\maketitle

\section{Introduction}
\label{sec:intro}

The phase structure of the cold dense nuclear matter has its significance as it relates to the chiral symmetry breaking, mediation of nuclear force, equation of state of compact star matter and compact star structures. Although it has been extensively investigated in the past several decades, there are still many uncertainties waiting for clarification, especially at density $\rho \gtrsim 2 \rho_0$ with 
$\rho_0 \approx 0.16$~fm$^{-3}$ being the saturation density~\cite{Fukushima:2010bq,Lattimer:2015nhk,Baym:2017whm,Ma:2019ery,Brandes:2023bob,Sorensen:2023zkk,Cai:2025nxn}. The theoretical origin of these uncertainties is attributed to the modeling of compact star matter which is governed by the strong interaction at low energies, as well as the inapplicability of lattice QCD simulations in this density region.

The duality between anti-de Sitter space and conformal field theory (AdS/CFT)~\cite{Maldacena:1997re,Gubser:1998bc,Witten:1998qj,Natsuume:2014sfa}, offers a novel tool to overcome (some of) these challenges. For strongly coupled QCD in four-dimensional spacetime, 
the AdS/CFT correspondence provides a non-perturbative approach that allows the 
exploration of low-energy QCD physics using analytical methods in the five-dimensional AdS (AdS$_5$) space based on the holographic principle with few parameters~\cite{Ammon:2015wua,Kim:2012ey}.  
Generally, there are two approaches in the construction of AdS/QCD models, the top-down approach which is obtained by reducing the (super-)string theories that give rise to the low-energy limit of QCD-like theories~\cite{Sakai:2004cn,Sakai:2005yt}, and bottom-up approach which is constructed by relating the basics of QCD models in four-dimensional spacetimes with the dual classical gravitational theory living in five-dimensional AdS (AdS$_5$)~\cite{Erlich:2005qh,DaRold:2005mxj}. In the literature, both approaches have been used in the study of nuclear matter~\cite{Hoyos:2021uff,Rougemont:2023gfz,Jarvinen:2023jbr}. Here we will consider the latter.

The core idea of AdS/QCD model from the bottom-up approach is to replicate the physical properties of QCD in four-dimensional spacetime in the AdS$_5$ spacetime by introducing additional fields, such as scalar 
and gauge fields. For example, the chiral symmetry breaking is implemented by introducing 
a five-dimensional scalar field $\Phi(z)$, where $z$ represents the holographic 
coordinate in AdS space and corresponds to the energy scale in four-dimensional 
QCD. The vacuum expectation value (VEV) of the scalar field, $\langle \Phi \rangle$, 
represents the magnitude of chiral condensates in QCD~\cite{Gherghetta:2009ac}. 
Boundary conditions at the 
infrared boundary enable effective descriptions of both chiral symmetry breaking 
and its restoration under finite density or temperature conditions~\cite{Chelabi:2015gpc, Ghoroku:2004sp}. Generally, the bottom-up AdS/QCD is divided into two categories according to how that $z$ coordinate is bounded, the hard-wall model~\cite{Erlich:2005qh, Cherman:2008eh, Kim:2009qs} 
and the soft-wall model~\cite{Karch:2006pv, Park:2011qq}. 
Each model has its own advantages and limitations: the hard-wall model features 
simple mathematical formulations and clear physical interpretations, while the 
soft-wall model excels at describing the linear trajectories of vector mesons 
\cite{Karch:2006pv}.

Finite density effect is typically introduced in AdS/QCD by adding a gauge field 
in the five-dimensional space, where the boundary value of this field corresponds to 
the chemical potential $\mu$~\cite{Colangelo:2011sr, Colangelo:2012jy}
. By analyzing the interaction between the gauge field and the 
scalar field, one can derive the relationship between chemical potential and baryon 
density $\rho$ ($\mu$-$\rho$ relation). In chiral phase transitions of nuclear matter, 
the vacuum expectation value (VEV) of the scalar field undergoes significant changes, which 
may manifest as first order (discontinuous) or continuous transitions. 
By combining chiral phase transitions and finite density 
effects, can effectively construct the EOS for nuclear matter, 
thus providing theoretical support for studying the mass-radius relation 
of neutron stars 
\cite{Bartolini:2022rkl, Kovensky:2021kzl, Jokela:2018ers}.

Previously, the hard-wall model has been used to study the QCD phase diagram, including the restoration of chiral symmetry and the temperature dependence of the critical chemical potential~\cite{Bartolini:2022rkl}. It was found that at zero temperature and finite density, chiral symmetry restores after the appearance of baryons. In this work, we employ a different homogeneous approximation~\cite{Elliot-Ripley:2016uwb}, which allows us to find a set of parameters such that the restoration of chiral symmetry occurs at high density/chemical potential in the baryonic phase. That is, the restoration of chiral symmetry decouples from the onset of baryons. In addition, we found that, in this new homogeneous approximation, the EOS is soften to coincide with the constraint from the empirical data.

The rest of this paper is organized follows: First, we introduce the 
action of the model, homogeneous approximation and the notation in Sec.~\ref{sec:model}. 
% Then, we present the Homogeneous Approximation in~\ref{subsec:homo_approx}. 
Next, in Sec.~\ref{sec:mattr}, we discuss the phase structure of nuclear matter through three different sets of parameters. 
The EOS for nuclear matter and the mass-radius relation of 
neutron stars are then discussed in Sec.~\ref{sec:neutron_stars}. 
Finally, we give the summary and discussion 
in Sec.~\ref{sec:summary}.

\section{The model and method}
\label{sec:model}

The anti-de Sitter (AdS) metric with an additional compactified extra dimension denoted by $z$ in AdS$_5$ is expressed as
\begin{equation}
ds^2 = e^{2A(z)} ( \eta_{\mu\nu} dx^{\mu} dx^{\nu} - dz^2 ),
\label{eq:metric_ads5}
\end{equation}
where $\eta_{\mu\nu}=\{1,-1,-1,-1\}$ is the Minkowski metric and $z$ is the holographic coordinate which is interpreted as the inverse energy scale. The function $A(z)$, known as the
warp factor, is given by
\be 
A(z) = \log \big( a(z) \big), \quad a(z)=\frac{L_0}{z} 
\ee
with $a(z)$ being the warp factor and $L_0$ as the radius of AdS curvature. In the present work, we choose $L_0=1$ for simplicity. We use $z_{uv}$ and $z_{ir}$ to represent UV (ultraviolet) and IR (infrared) 
cutoff respectively, and compactify the holographic coordinate $z$ within the range $z_{uv} \leq z \leq z_{ir}$~\cite{Randall:1999ee}.

In the following, we will consider the scalar field $\Phi$ which transforms under local $U(2)_L \times U(2)_R$ symmetry. Its VEV results in the breaking of chiral symmetry. In addition, corresponding to the local $U(2)_L \times U(2)_R$ symmetry, we will include the gauge fields  $\mathcal{L}_M$ and $\mathcal{R}_M$ in the bulk, which are dual to the left- and right-handed quark currents respectively. 

\subsection{The action}
\label{subsec:action}

The minimal 5D action can be divided into four parts~\cite{DaRold:2005mxj, Domenech:2010aq, DaRold:2005vr}:
\be
S = S_{g} + S_{CS} + S_{\Phi} + S_{IR}.
    \label{eq:action_ads}
\ee

The pure gauge part of the action is given by:
\be
S_{g} & = &{} - \int d^{5}x \frac{\sqrt{g}}{2 g_{5}^{2}} \bigg\{ 
                \operatorname{Tr} \Big( \mathcal{L}_{MN} \mathcal{L}^{MN} \Big) 
                + \big( \mathcal{R} \leftrightarrow \mathcal{L} \big) \bigg\}
%            \nonumber\\
%& = &{} - \int d^{4}x \int_{z_{uv}}^{z_{ir}} dz \frac{1}{2} \frac{1}{g_{5}^{2}} \sqrt{g} 
%            \bigg\{ \operatorname{Tr} \Big( L_{MN}L^{MN} \Big)
%            \nonumber\\ 
%& & \qquad\qquad\qquad\qquad\qquad\quad{} + \frac{1}{2}\hat{L}_{MN}\hat{L}^{MN} + \big( R \leftrightarrow L \big) \bigg\}
\label{eq:action_gauge}
\ee
where $g$ is the absolute value of the determinant of the metric (\ref{eq:metric_ads5}), and the gauge coupling 
$g_5^2 = 12\pi^2/N_c$ is determined from the calculation of the two-point function of vector current 
\cite{Erlich:2005qh, DaRold:2005mxj}. The $5D$ gauge field ${\cal L}_M$ can be decomposed in terms of the $SU(2)$ and $U(1)$ components as
\be
{\cal L}_M = {L_M} + \frac{\mathbbm{1}}{2}{\hat L_M}, \quad
    L_M = \frac{{{\tau ^a}}}{2} L_M^a, 
\ee
where ${L_M}$ and ${\hat{L}_M}$ 
are the $ SU(2) $ part and $U(1)$ part of the $U(2)$ field $\mathcal{L}_M$, respectively, and $ \operatorname{Tr} \big( \tau^a \tau^b \big) = 2 \delta^{ab}$ with $\tau^a$ are the Pauli matrices. 
%${L_M}$ and ${\hat{L}_M}$ are the $ SU(2) $ part and $U(1)$ part of the $U(2)$ field $\mathcal{L}_M$ respectively. 
${\cal L}_{MN}$ is the field strength tensor of gauge field ${\cal L}_M$ with
\be 
{\cal L}_{MN} & = & \partial_M {\cal L}_N - \partial_N {\cal L}_M - i[{\cal L}_M, {\cal L}_N].
\ee
Similar arguments apply to the gauge field $\mathcal{R}_M$.

In terms of the $U(2)$ gauge fields, one can write the Chern-Simons (CS) action as~\cite{Domenech:2010aq, Bartolini:2022rkl}:
\be
S_{CS} & = & \frac{N_c}{32 \pi^2}\int d^5 x \varepsilon^{MNOPQ} 
                \bigg[ \frac{1}{2} \hat{L}_M  \operatorname{Tr} \left( L_{NO}L_{PQ} \right) \nonumber\\
& & \hspace{5 mm} + \frac{1}{12} \hat{L}_M \hat{L}_{NO} \hat{L}_{PQ} 
% \nonumber\\
- \left( L \to R, \hat{L} \to \hat{R} \right) \bigg],
\label{eq:action_cs}
\ee
where $N_c$ is the number of colors.

In addition to the two $U(2)$ gauge fields, the flavor content of the model is also given by a bi-fundamental complex scalar $\Phi(x,z)$ which is due to the order of the chiral symmetry breaking in QCD. The action of the scalar field part is
\be
S_{\Phi} & = & \int d^5 x \frac{1}{g_{5}^{2}} \sqrt{g} \bigg\{
        \operatorname{Tr} \left[ \big(D_M \Phi \big)^{\dagger} D^M \Phi \right]
        \nonumber \\
& &{} \qquad\qquad\qquad\qquad - m_5^2(z) \operatorname{Tr} \left[ \Phi^{\dagger} \Phi \right] \bigg\},
    \label{eq:action_phi}
\ee
where $ D_M\Phi=\partial_M\Phi-i\mathcal{L}_M\Phi+i\Phi\mathcal{R}_M $ is the covariant derivative. The 5D mass $m_5^2 = -3$ is determined by the AdS/CFT dictionary. In principle, there should be corrections to 5D mass originated from the anomalous dimension of quark mass operator which is constrained by the mass splitting of chiral partners~\cite{Fang:2016nfj, Fang:2018axm}. 
% \sout{Here, without loss of generality, we will not consider this correction but taking $m_5^2 = -3$.}
% \cmb{In this work, we do not study the mass splitting between vector and axial-vector mesons, so we set $m_5^2 = -3$, which neglects the corrections present in the soft-wall model~\cite{Fang:2016nfj, Fang:2018axm}; this is also a common practice when working with the hard-wall model.}
Here, we will not consider this correction but simply take $m_5^2 = -3$.

To realize the spontaneous chiral symmetry breaking, we include an action at the IR boundary~\cite{DaRold:2005vr, Evans:2004ia, DaRold:2005mxj}:
\be
\label{eq:action_ir1}
S_{IR} & = &{} - {e^{4A(z)}} \int d^4 x V(\Phi) \Big|_{z_{ir}}, \\
\label{eq:action_ir2}
V(\Phi) & = &{} -\frac{1}{2} m_{1} \operatorname{Tr} \big(\Phi^{\dagger} \Phi \big) 
+ m_{2}\operatorname{Tr} \big( \Phi^{\dagger} \Phi \big)^2, \nonumber
\ee
where $m_1$ and $m_2$ are free parameters. $V(\Phi)$ is a potential for the scalar field $\Phi$ which allows for a non-zero expectation value for $\Phi$ on the IR-boundary, even in the 
chiral limit $M_q = 0$, i.e., a non-zero expectation value $\Phi |_{z_{ir}}$ 
corresponding to the spontaneous breaking of chiral symmetry. $\Phi$ has VEV
\be
\langle \Phi \rangle = \frac{\mathbbm{1}}{2} \omega_0(z), 
\ee
According to the AdS/CFT dictionary~\cite{Erlich:2005qh}, the VEV of scalar field $\Phi$ 
has the following behavior at zero baryon number density 
%\cmh{(however, the situation for the scenario at finite density will differ):}
\be
\omega_0(z) = M_q z + \Sigma z^3 + \cdots,
\label{eq:omega_sigma}
\ee
where $M_q$ and $\Sigma$ are dual to the current quark mass and the chiral condensate, respectively.
%\cmh{which defined in~\cite{Erlich:2005qh} is $1/g_5$ times of ours.}
%\cmb{the define of $\Sigma$ in Ref.~\cite{Erlich:2005qh} is $1/g_5$ times of ours. }
It should be noted that, at finite density, the relation between $\omega_0$ and $\Sigma$ is no longer the simple expression shown in Eq.~\eqref{eq:omega_sigma} due to the density dependence of the EOM of $\langle \Phi \rangle$, making the solution of $\Sigma$ complicated. In the following discussions, we will use $\omega_{ir}$ (the value of $\omega_0(z_{ir})$, defined in Eq.~\eqref{eq:bc_omega}) to characterize the magnitude of the chiral condensate $\Sigma$.

For the indices, we use the following conventions: Capital Latin letters $M, N, \cdots$ 
run over 4D spacetime $(t,x)$ and the holographic dimension $z$; Greek letters $\mu, \nu, \dots$ run over 4D spacetime $(t,x)$; Lowercase Latin letters $i, j, \dots$ run 
over the 3D spatial indices $(x)$; The number $0$ and the letter $z$ specifically refer to the time coordinate 
$t$ and the holographic coordinate $z$, respectively.

\subsection{Homogeneous approximation}
\label{subsec:homo_approx}

In the present Holographic model approach to nuclear matter, we will not include the fermion field as an explicit degree of freedom~\cite{Kim:2007xi}, but instead interpret baryons as solitonic configurations in the flavor gauge fields~\cite{Ghoroku:2012am,Ghoroku:2013gja}. To study nuclear matter properties, we use the homogeneous approximation from kinky holographic nuclear matter~
\cite{Elliot-Ripley:2016uwb}, which assumes a static and spatially homogeneous configuration. 
This means that the gauge fields $L_M, R_M, \hat{L}_M, \hat{R}_M$ and the scalar field $\Phi$ only depend on 
the holographic coordinate $z$, specifically, as:
\begin{subequations}
\be
& & L_i = -R_i = \beta \varphi(z) \frac{\tau_i}{2}, \quad L_z = R_z = 0 ,
    \label{eq:homogeneous_01} \\
& & {\hat L}_0 = {\hat R}_0 = \hat {c}_0(z), \quad {\hat L}_{0z} = {\hat R}_{0z} =  - \hat {c}_0 ^{\prime} ,
    \label{eq:homogeneous_02} \\
& & L_{iz} = {} - {R_{iz}} = -\beta \varphi ^{\prime} \frac{1}{2}{\tau _i},
    \label{eq:homogeneous_03} \\
& & L_{ij} = R_{ij} = \frac{1}{2}{\beta ^2}\varphi (\varphi-1 ){\varepsilon _{ijk}}{\tau _k} , \label{eq:homogeneous_04}
\ee
\end{subequations}
where $\beta$ is a parameter associated with the baryon number density $\rho$, which will be seen later. $\varphi$ is a real field in order to generate non-trivial topological charge.
$\hat{c}_0 (z)$ is a field associated with the baryon chemical potential $\mu$. Note that Eq.~(\ref{eq:homogeneous_04}) is slightly different from the approximation used 
in Refs.~\cite{Rozali:2007rx, Kim:2007vd}. This is because of the inclusion of the equation of motion (EOM) for the pure gauge field $U(x)$~\cite{Elliot-Ripley:2016uwb, Sutcliffe:2010et}, as shown in App.~\ref{app}, which was previously neglected. Later we will see that, by including the effects of the pure gauge fields, we can find a suitable set of parameters such that a chiral phase transition occurs in the baryonic phase at zero temperature, and this inclusion also improves the EOS to align with empirical constraints.

Substituting Eqs.~(\ref{eq:homogeneous_01}-\ref{eq:homogeneous_04}) into the action (\ref{eq:action_ads}), we obtain the simplified action as follows:
%\begin{subequations}
\be
S_g & = & \int d^5x \frac{1}{g_5^2} \left\{ -3e^{A}\beta^4\varphi^2 ( 1-\varphi )^2
        - 3e^{A}\beta^2\varphi^{\prime 2}
        + e^{A}\hat{c}^{\prime 2}_0 \right\}, \nonumber\\
S_{CS} & = & \int d^5x \frac{3N_c}{8\pi^2} \beta^3 \hat{c}_0 \varphi^{\prime} \varphi (1-\varphi) , \nonumber\\
S_{\Phi} & = & \int d^{5}x\frac{1}{g_{5}^{2}} e^{3A} \left\{-\frac{1}{2} \omega_{0}^{\prime}{}^{2}
        -\frac{3}{2}\varphi^{2}\beta^{2} \omega_{0}^{2}
        -\frac{1}{2} e^{2A}m_{5}^{2} \omega_{0}^{2} \right\}, \nonumber\\ 
S_{IR} & =&{} -e^{4A} \int d^4x \left( -\frac{1}{4} m_1 \omega_0^2 + \frac{1}{8} m_2 \omega_0^4 \right),
\label{eq:action_homo}
\ee
%\end{subequations}
where $P^\prime \equiv \frac{\partial P}{\partial z}$ for $P \in \{ \varphi, \omega_0 \}$.

\section{Phase structure of baryonic matter}
\label{sec:mattr}

Now, we are ready to study baryonic matter properties using the theoretical framework discussed above and then apply the obtained EOS to study the neutron star structures.

\subsection{Baryonic phase at finite density}
\label{subsec:baryonic_phase}

For convenience, we define the following dimensionless fields and parameters with tildes: 
\be
& & \widetilde{z}_{ir} = 1, \ 
    \widetilde{\beta} = \beta z_{ir}, \ 
    \widetilde{\hat{c}}_0 = \hat{c}_0 z_{ir}, \nonumber\\ 
& & \widetilde{M}_q = M_q z_{ir}, \ 
    \widetilde{\Sigma} = \Sigma z_{ir}^3.
\ee
That is, $ z_{ir} $ becomes a rescaling factor. Unless otherwise specified, we use symbols without tildes to represent these 
dimensionless fields and parameters, subsequently.

The energy and grand potential $\Omega$ defined in Eq.~\eqref{eq:eos} can be obtained by solving the EOMs:
\begin{subequations}
\be
{\varphi ^{\prime \prime }} & = &{}  - {A^\prime }{\varphi ^\prime } + {\beta ^2}(\varphi  
        - 3{\varphi ^2} + 2{\varphi ^3}) + \frac{1}{2}{e^{2A}}\varphi \omega _0^2 \nonumber\\
& &{} + \frac{1}{2}\frac{{{N_c}}}{{8{\pi ^2}}}\beta 
        \hat c_0^\prime \varphi (1 - \varphi )g_5^2{e^{ - A}},\label{eq:diffeq_phi}
        \\
\hat{c}_{0}^{\prime\prime} & = &{} -A^{\prime}\hat{c}_{0}^{\prime}
        +\frac{1}{2}\frac{3N_{c}}{8\pi^{2}}g_{5}^{2}e^{-A}\beta^{3}\varphi^{\prime}\varphi(1-\varphi),
        \label{eq:diffeq_c0}\\ 
\omega _0^{\prime \prime } & = &{} - 3{A^\prime }\omega _0^\prime  + 3{\varphi ^2}{\beta ^2}{\omega _0} + m_5^2{\omega _0}{e^{2A}},\label{eq:diffeq_omega}
\ee
\end{subequations}
subject to the following boundary conditions (BCs)
\begin{subequations}
\be
\label{eq:bc_varphi} 
& & \varphi(z_{uv}) = 0, \quad \varphi(z_{ir}) = 1, \\
\label{eq:bc_c0}
& & \hat{c}_0(z_{uv}) = \mu_c, \quad \hat{c}_0^{\prime}(z_{ir}) = 0, \\
\label{eq:bc_omega}
& & \omega_0(z_{uv}) = 0, \quad \omega_0(z_{ir}) = \omega_{ir},
\ee
\end{subequations}
where $\mu_c$ and $\omega_{ir}$ are free parameters. $\mu_c$ is related to chemical potential which is chosen as $\mu_c = 0$ in matter-free space, 
and there is a relation between $\omega_{ir}$ and the chiral condensate $\Sigma$ as shown in Eq. \eqref{eq:omega_sigma} at zero baryon number density.

From the CS term, one can express the baryon number, the topological charge, as 
\be
B %& = & \frac{1}{{32{\pi ^2}}}\int {{d^3}x} \int {dz} {\varepsilon ^{MNPQ}}
%        \operatorname{Tr} \Big[ {L_{MN}}{L_{PQ}} - {R_{MN}}{R_{PQ}} \Big]
%    \nonumber \\
& = & \frac{{3V}}{{4{\pi ^2}}}\int {dz}  {\beta ^3}\varphi ^{\prime} \varphi (1 - \varphi ) 
    = V \frac{{\beta ^3}}{{8{\pi ^2}}} ,
\label{eq:baryon_number}
\ee
where $V$ is the volume of the system. To obtain the last equality, we used the boundary condition of $ \varphi(z) $ (\ref{eq:bc_varphi}). Then, the baryon number density is expressed as:
\be
\rho = \frac{B}{V} = \frac{\beta^3}{8 \pi^2}.
\label{eq:rel_rho}
\ee
In other words, up to a normalization factor, the parameter $\beta$ is equivalent to the baryon number density $\rho$. 

% In addition, from the CS term, we have
% \be
% S_{CS} & = & \int d^5x \frac{3N_c}{8\pi^2} \beta^3 \varphi^{\prime} \varphi (1-\varphi) \hat{c}_0(z) \nonumber\\
% & = & \int {{d^5}} x\frac{{3{N_c}}}{{8{\pi ^2}}}{\beta ^3}{\varphi ^\prime }\varphi (1 - \varphi )({\mu _{c0}} + {\mu _c})
%     \nonumber\\
% & = & {S_{CS}}{\big |_{{\mu _c} = 0}} + \frac{N_c}{2}{\mu _c}\rho ,
%     \label{eq:mu_muc}
% \ee
% where $\mu_{c0}(z) \equiv \hat{c}_0(z) \big|_{\mu_c=0}$ satisfying $\mu_{c0}(z_{uv}) = 0$. So, we obtain the baryon chemical potential as
% \be
% \mu = \frac{N_c}{2} \mu_c. 
% \label{eq:mu_define}
% \ee

\begin{figure}[tb]
    \centering
    \includegraphics[width=0.8 \linewidth]{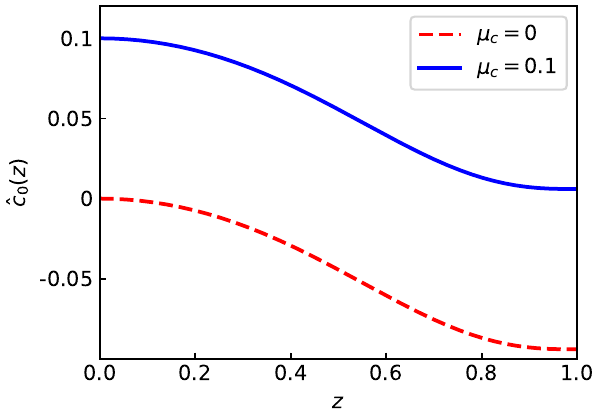}
    \caption{The image obtained under different $\mu_c$, with all other parameters fixed. The solution when $\mu_c \ne 0$ is equivalent to the solution when $\mu_c = 0$, but shifted by constant $\mu_c$.}
    \label{fig:difft_muc}
\end{figure}

The definition of the chemical potential can be understood alternatively as follows. By integrating out Eq.~(\ref{eq:diffeq_c0}), one obtains 
\begin{equation}
\hat{c}_0(z) = \mu_c + \frac{3}{2} N_c g_5^2 \rho \int_{z_{uv}}^z dz e^{-A} \left( \frac{\varphi^2}{2} - \frac{\varphi^3}{3} - \frac{1}{6} \right).
\end{equation}
This equation shows that the UV boundary condition only shifts the profile of \( \hat{c}_0(z) \), as confirmed numerically in Fig.~\ref{fig:difft_muc}. After substituting this formula into Eq.~(\ref{eq:diffeq_phi}) and Eq.~(\ref{eq:diffeq_omega}), one can solve them with respect to the BCs (\ref{eq:bc_varphi}) and (\ref{eq:bc_omega}). Then, the on-shell action is expressed as
\begin{equation}
S_{OS} = V \left( S(\beta, \omega_{ir}) + \frac{N_c}{2} \mu_c \rho \right).
\end{equation}
In this action, the only term affected by the UV boundary condition of \( \hat{c}_0 \), namely \( \mu_c \), is \( S_{CS} \). This term can be separated and expressed as the \( \mu_c \rho \) term in the equation above, while the other terms are not influenced by \( \mu_c \) and are written as \( S(\beta, \omega_{ir}) \). Then the grand potential, obtained through the holographic dictionary at zero temperature, is expressed as:
\begin{equation}
\Omega = - \frac{S_{OS}}{V} = -S(\beta, \omega_{ir}) - \frac{N_c}{2} \mu_c \rho.
\end{equation}
Therefore, it is natural to define the baryon chemical potential as:
\begin{equation}
\mu = \frac{N_c}{2} \mu_c.
\label{eq:mu_define}
\end{equation}
And, thermodynamic self-consistency requires that:
\begin{equation}
\frac{\partial \Omega}{\partial \beta} = \frac{\partial \Omega}{\partial \omega_{ir}} = 0.
\end{equation}

In our model, there are six parameters: $z_{ir}$, $m_1$, $m_2$, $\beta$, $\mu_c$ and $\omega_{ir}$. These parameters can be determined as follows:
\begin{enumerate}
    \item The parameters $\beta$, $\omega_{ir}$ can be determined by minimizing the grand potential $\Omega$ at every fixed $m_1$, $m_2$ and $\mu_c$(or equivalently, chemical potential $\mu$). 

    \item The parameter $z_{ir}$ is determined by $\mu_0$ or $\rho_0$ with $\mu_0$ denoting the chemical potential at the onset of baryons and $\rho_0$ the corresponding baryon number density. 
Due to the rescaling by $ z_{{ir}} $, we can estimate the value of $ z_{{ir}} $ by using the empirical values of the baryon mass $m_n = 938$~MeV , binding energy $ B.E.\approx -16$~MeV and $\rho_0 \approx 0.16$~fm$^{-3}$. 
This is different from the procedure in vacuum where $z_{ir}$ is estimated by using the mass of $\rho$ meson~\cite{DaRold:2005mxj,Erlich:2005qh}.

\item For convenience, we define a new parameter $D = \mu_{c0}/\beta_0 $. The relationship between $\mu_{c0}$, $\beta_0$ and $\mu_0$, $\rho_0$ is given in Eq.  \eqref{eq:mu_define} and \eqref{eq:rel_rho}.
Then parameters $m_1$ and $m_2$ are estimated by fitting the value of $D$ in vacuum 
and chiral condensation $\Sigma$ (or $\omega_{ir} |_{\rho=0}$). 
\end{enumerate}

To reveal the properties of nuclear matter, along the above procedure, we choose three sets of 
parameters shown in Tab.~\ref{tab:parameters}: 
The set without $\Phi$ with $m_1=0$ and $m_2=0$ 
means that the effects of the scalar field are ignored. In this case we fix $D=1.41$. In set A, we estimated $m_1$ and $m_2$ by fitting the values of $\omega_{ir} |_{\rho=0} = 4\sqrt{2}$ and $D=1.41$ (equivalent to $\xi=4$ in Ref.~\cite{DaRold:2005vr} which is estimated by fitting the meson spectrum and it is consistent with the value used in Ref.~\cite{Bartolini:2022rkl}). In set B, $m_1$ and $m_2$ are estimated by fitting the values of $\Sigma^{1/3}/z_{ir}\approx200$~MeV and $D\approx1.45$.

\begin{table}[htbp]
    \centering
    \begin{tabular}{lll}
        \hline\hline
        parameter sets  &  \hspace{5 mm} $m_1$ & \hspace{5 mm} $m_2$ \\
        \hline
         without $\Phi$ &  \hspace{5 mm} 0     & \hspace{5 mm} 0 \\
         set A          &  \hspace{5 mm} 0.283 & \hspace{5 mm} $4.12 \times 10^{-3}$ \\
         set B          &  \hspace{5 mm} 0.510 & \hspace{5 mm} $4.63 \times 10^{-2}$ \\
         \hline\hline
    \end{tabular}
    \caption{
        Parameter sets used in this work. 
    }
    \label{tab:parameters}
\end{table}

\subsection{Phase structure of baryonic matter}
\label{subsec:phase}

\begin{figure}[b]
    \centering
    \includegraphics[width=0.8 \linewidth]{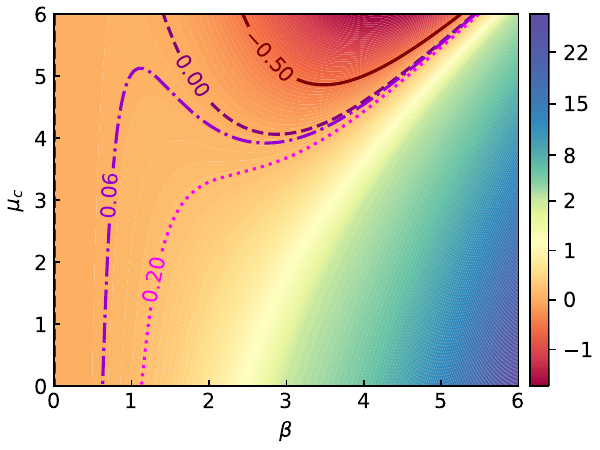}
    \caption{
        Contour plot of the grand potential $\Omega$ without including 
    scalar field $\Phi$. 
    }
    \label{fig:without_x_contourf}
\end{figure}

\begin{figure}[tbp]
    \centering
    \includegraphics[width=0.8 \linewidth]{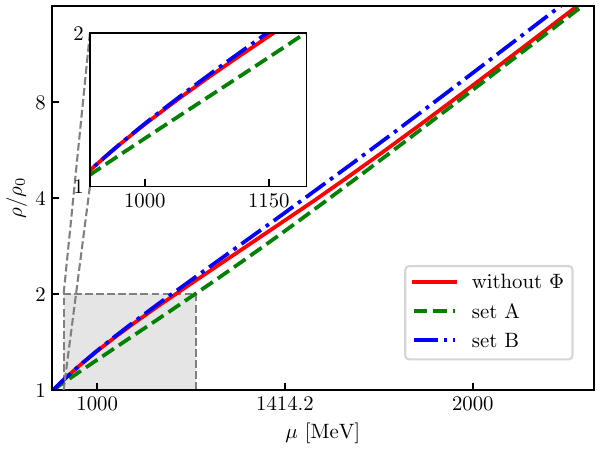}
    \caption{The log-log plot of baryon chemical potential $\mu$ versus $\rho/\rho_0$.}
    \label{fig:mu_rho}
\end{figure}

First, let us study the case without including the scalar field, as was done in Ref.~\cite{Braga:2024nnj}. In this case, we can determine the 
relationship between the baryon chemical potential (parameter $\mu_c$ or $\mu$) and the 
baryon number density (parameter $\beta$ or $\rho$) by minimizing 
the grand potential $\Omega$. The results are shown in 
Figs.~\ref{fig:without_x_contourf} and~\ref{fig:mu_rho}. 

In Fig.~\ref{fig:without_x_contourf}, we plot the contour of the grand potential $\Omega$ as a function of $\mu_c$ and $\beta$. Additionally, 
we provide four contour lines for different values of $\Omega$ to help the readers 
understand the results better. 
The brown-solid line represents the contour for $\Omega = -0.5$, the purple-dashed line 
represents $\Omega = 0$, the dark-violet dot-dashed line represents $\Omega = 0.06$, 
and the magenta-dotted line represents $\Omega = 0.2$. 
It should be noted that there are two lines for $\Omega = 0$: 
the one indicated by the purple-dashed line and the other is along the axis with $\beta = 0$. 

From Fig.~\ref{fig:without_x_contourf}, one can observe that, when the chemical potential $\mu_c$ is relatively small, the grand potential has only one global minimum located at the axis $\beta = 0$; As $\mu_c$ increases, the grand potential begins to 
exhibit local extrema in regions with $\beta \ne 0$. For example, the contour of 
the dark-violet dot-dashed line shows a local minimum at the point 
($\beta \approx 2.6$, $\mu_c \approx 3.9$).
When the chemical potential is sufficiently large, the local minimum
becomes the new global one, such as the contours of $\Omega<0$; 
The lowest point of the purple-dashed line corresponds to the point of 
the nuclear saturation density $\rho_0=0.16$~fm$^{-3}$, 
indicating the emergence of baryons at this point; 
As the chemical potential $\mu_c$ continues to increase, the density $\beta$ 
corresponding 
to the minimum also increases, thus the relationship between the 
chemical potential $\mu$ and the baryon number density $\rho$ could be 
established, as shown by the red-solid line in Fig.~\ref{fig:mu_rho}. 

In Fig.~\ref{fig:mu_rho}, we present the correlation between 
the chemical potential $\mu$ and normalized baryon number 
density $\rho/\rho_0$. 
The red-solid line represents the results without including 
scalar field $\Phi$. The green-dashed line corresponds to the 
results fitted with $\omega_{ir}=4\sqrt{2}$ and $D=1.41$ (set A). The blue-dash-dotted line represents 
the fitted results of the chiral condensation $\Sigma^{1/3}/z_{ir}\approx 200$~MeV and $D\approx 1.45$ (set B). 
We can see that the green-dashed line and red-solid 
line gradually converge, moreover, at high densities, the three curves exhibit a trend toward convergence. Reflecting that the three sets of parameters exhibit the same behavior at high density. The same behavior is also observed in the EOS shown later in Fig.~\ref{fig:e_p}.

We next turn to the speed of sound in the baryonic
matter phase which is defined as:
\be
v_s^2 = \frac{\partial P}{\partial \mathcal{E}} = \frac{\rho}{\mu} \frac{\partial \mu}{\partial \rho},
\label{eq:vs2_0}
\ee
where $P$ is the pressure and $\mathcal{E}$ denotes the energy density and 
\be
\rho & = & \frac{\partial P}{\partial\mu},\quad \mu=\frac{\partial\mathcal{E}}{\partial\rho} .
\label{eq:rho_mu_t0}
\ee

\begin{figure}[tb]
    \centering
    \includegraphics[width=0.8 \linewidth]{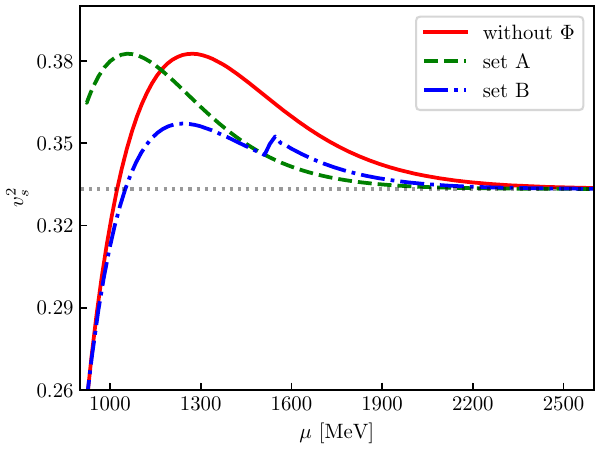}
    \caption{The sound speed plots under three different sets of parameters.}
    \label{fig:mu_vs2}
\end{figure}

\begin{figure}[tb]
    \centering
    \includegraphics[width=0.8\linewidth]{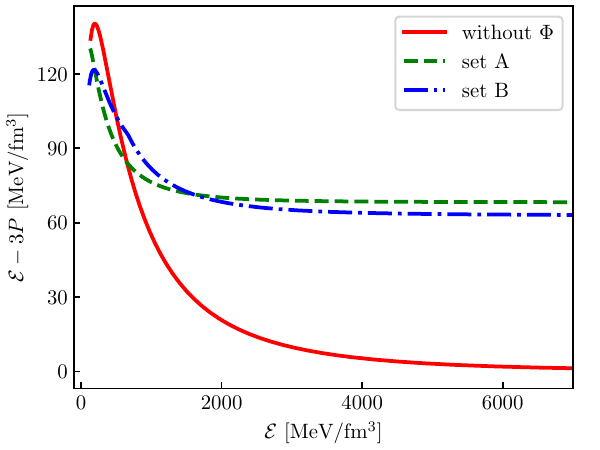}
    \caption{The value of $\mathcal{E}-3P$ from three sets of parameters. }
    \label{fig:e-3p}
\end{figure}

The speed of sound as a function of chemical potential is plotted in Fig.~\ref{fig:mu_vs2}. One can see that it exceeds $1/3$---the conformal limit---for all three parameter sets at intermediate densities, 
which may satisfy the empirical constraints from neutron star observations
\cite{Bedaque:2014sqa, Kojo:2014rca, Tews:2018kmu}. 
We observe that if no chiral phase transition occurs (as shown in Fig.~\ref{fig:mu_oir}), the cases 
with and without scalar field $\Phi$ yield the 
same maximum value for the speed of sound, shown in the red and green lines of Fig.~\ref{fig:mu_vs2}. 
However, the existence 
of the chiral phase transition (as shown in Fig.~\ref{fig:mu_oir}) reduces the maximum value of the 
speed of sound and induces a discontinuous jump at the transition point, shown in the blue line of Fig.~\ref{fig:mu_vs2}. It is interesting to note that the speed of sound converges to the conformal limit for all the three parameter sets at densities $\approx 6 \rho_0$, a density relevant to the cores of massive stars. This aligns with the pseudoconformal structure of compact star matter~\cite{Ma:2018xjw,Ma:2018jze} since, as shown in Fig.~\ref{fig:e-3p}, the trace of the energy-momentum tensor, does not vanish. Another interesting point is that, in all three cases, the speed of sound exhibits a peak, irrelevant to the existence of phase transition. This may be related to how the scale symmetry manifests in nuclear matter~\cite{Fujimoto:2022ohj,Zhang:2024sju} which deserves to study in the future in the framework of AdS/QCD models.

\subsection{Chiral phase transition in baryonic phase}
\label{subsec:chiralphase}

We next consider the behavior of chiral symmetry breaking---a significant quantity in hadron physics---in nuclear matter. In the literature, using holographic QCD models, some investigations have been performed in the thermal phase and quark-gluon plasma (QGP) phase~\cite{Bartz:2024dgd, Bartz:2016ufc, Cherman:2008eh}. Here, we focus on the baryonic phase at zero temperature.

\begin{figure*}[htbp]
    \centering
    \begin{subfigure}[b]{0.32\linewidth}
        \centering
        \begin{overpic}[width=\linewidth]{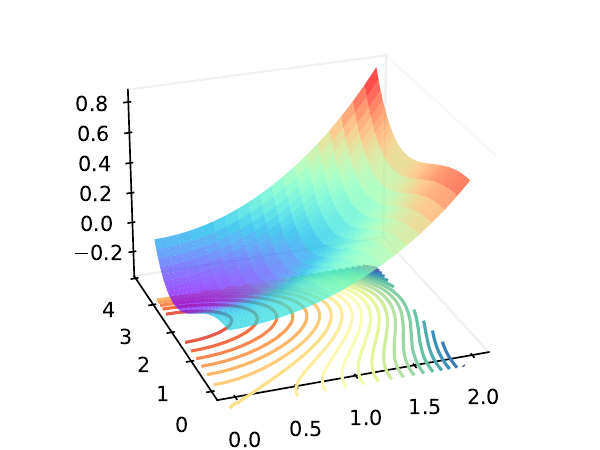}
            \put(7, 45){$\Omega$}
            \put(12, 12){$\omega_{ir}$}
            \put(70, 0.1){$\beta$}
        \end{overpic}
        \caption{$\mu_c=0$}
        \label{fig:with_x_muc_0}
    \end{subfigure}
    \hfill
    \begin{subfigure}[b]{0.32\linewidth}
        \centering
        \begin{overpic}[width=\linewidth]{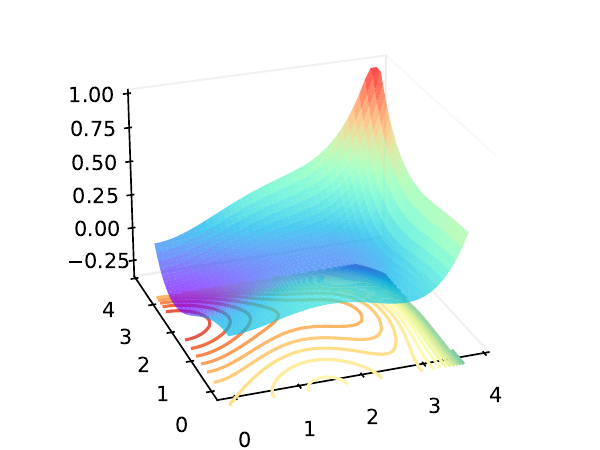}
            \put(7, 45){$\Omega$}
            \put(12, 12){$\omega_{ir}$}
            \put(70, 0.1){$\beta$}
        \end{overpic}
        \caption{$\mu_c=4$}
        \label{fig:with_x_muc_4}
    \end{subfigure}
    \hfill
    \begin{subfigure}[b]{0.32\linewidth}
        \centering
        \begin{overpic}[width=\linewidth]{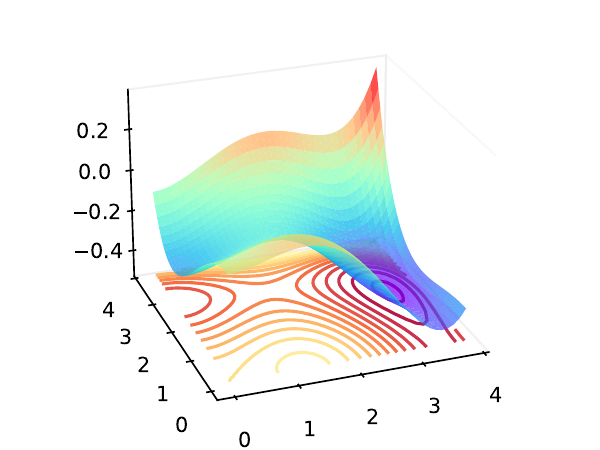}
            \put(7, 45){$\Omega$}
            \put(12, 12){$\omega_{ir}$}
            \put(70, 0.1){$\beta$}
        \end{overpic}
        \caption{$\mu_c=4.77$}
        \label{fig:with_x_muc_477}
    \end{subfigure}
    \caption{The grand potential $\Omega$ with scalar 
        field $\Phi$ at fixed $\mu_c$ in parameter set B. 
        The colored surfaces in three figures are 3D plots of $\Omega$. 
        Below the surfaces are the contour lines of $\Omega$. 
        Because of $\Omega_{min} \neq 0$ at $\mu_c=0$, one should 
        subtract this zero-point energy. For the colored surface, a deeper blue indicates a smaller value of $\Omega$.
    }
    \label{fig:with_x_muc}
\end{figure*}

\begin{figure}[htb]
    \centering
    \includegraphics[width=0.8\linewidth]{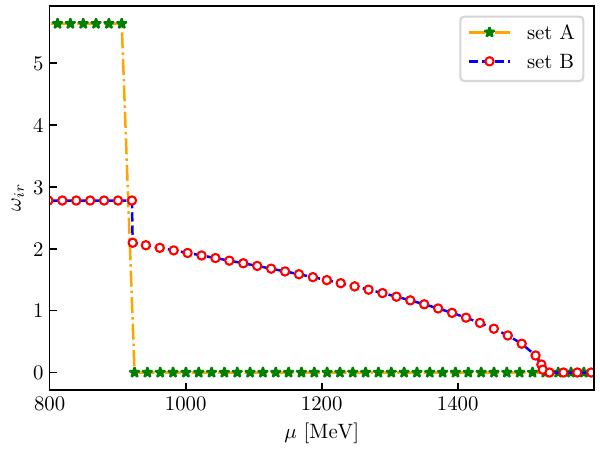}
    \caption{Plot of $\omega_{{ir}}$ as a function of $\mu$.
    %The green stars represent the values of $\omega_{ir}$ in set A, the red circles represent the values in set B.
    % When $\mu_c < 4.5$, the value of $\omega_{{ir}}$ equals $\omega_{{ir}}|_{\rho=0}$; $\mu_c \approx 4.5$ corresponds to the chemical potential at which baryons appear; afterwards, as $\mu_c$ increases, $\omega_{{ir}}$ gradually decreases; until $\mu_c \approx 7.5$, $\omega_{{ir}}$ decreases to zero.
    When $\mu < 922$~MeV, $\omega_{ir}$ is constant and equal to its vacuum value. When $\mu > 922$~MeV, $\omega_{ir}$ remains $0$ in set A; While in set B, $\omega_{ir}$ slowly decreases with $\mu$, until it vanishes at $\mu \approx 1535$~MeV.}
    \label{fig:mu_oir}
\end{figure}

In the hard-wall holographic model, the chiral condensation 
$\Sigma$  at zero density is determined by the scalar field term~\eqref{eq:action_phi} and the 
infrared action~\eqref{eq:action_ir1}. At finite density, the existence of the chiral phase 
transition depends on the choice of parameters. In set A, the chiral condensate is already zero at nuclear 
saturation density ($\rho_0$ or $\beta_0$), i.e., chiral symmetry breaking 
only occurs before the onset of baryons. This is consistent with the results 
in Ref.~\cite{Bartolini:2022rkl}.
In set B, the chiral condensation~\footnote{At zero density, chiral 
    condensation $\Sigma$ and $\omega_{ir}$ are equivalent. At finite density, although 
    they are not strictly equivalent, $\omega_{ir}$ can still be considered 
    as an order parameter for whether the chiral condensation is zero.
} 
undergoes a discontinuous 
jump (first-order phase transition) at nuclear saturation 
density, and then decreases with increasing chemical potential 
until it reaches zero, shown in Figs.~\ref{fig:with_x_muc} and~\ref{fig:mu_oir}. So the chiral transition in set B is second-order phase transition happened at $\mu \approx 1535$~MeV.

In Fig.~\ref{fig:with_x_muc}, we present the grand potential $\Omega$ as 
a function of baryon number density $\beta$ and the infrared 
value $\omega_{ir}$ of the scalar field VEV $\omega_0$ at different chemical potentials. The case without scalar field corresponds to 
the $\omega_{ir} = 0$ axis. 
Due to the presence of the infrared action, a local minimum appears 
along the $\omega_{ir}$-axis at $\omega_{ir} \neq 0$ (corresponding to 
the global minimum point in Fig.~\ref{fig:with_x_muc_0}), 
and this minimum does not change with chemical potential $\mu_c$. 
When $\mu_c$ is relatively small, the grand potential 
has only one global minimum at $\omega_{ir} \neq 0$, while at the origin 
($\omega_{ir} = 0, \beta=0$), it forms a saddle 
point, shown in Fig.~\ref{fig:with_x_muc_0}. As $\mu_c$ increases, 
the grand potential starts to exhibit local extremum in regions where both $\beta \neq 0$ and $\omega_{ir} \neq 0$, shown in Fig.~\ref{fig:with_x_muc_4}. 
When the chemical potential $\mu = \frac{N_c}{2} \mu_c = \mu_0$, there will be two global extrema: one is at ($\beta=0$, $\omega_{ir} \neq 0$); another one is at ($\beta \neq 0$, $\omega_{ir} \neq 0$), which is the nuclear saturation density point $\rho_0$. As $\mu_c$ continues to increase, the nuclear saturation density point becomes the new unique global minimum, shown in Fig.~\ref{fig:with_x_muc_477}. The contour lines in Fig.~\ref{fig:with_x_muc} illustrate the same conclusions. Eventually, the chiral condensate decreases to zero, and chiral symmetry is fully restored.

To illustrate the evolution of chiral symmetry with the baryon chemical potential, we show the behavior of $\omega_{{ir}}$ as a function of $\mu$ in Fig.~\ref{fig:mu_oir}. At $\mu=922$~MeV ($\mu_c = 4.9$ and $4.5$ in set A and B, respectively), there is a discontinuity in the figure, which arises from the global minimum of  $\Omega$ shifting from the vacuum phase to the baryonic phase. 
At $\mu=1535$~MeV ($\mu_c \approx 7.5$ in set B), there is a point where the first-order derivative is discontinuous, which corresponds to the critical point of the chiral phase transition at zero temperature and finite density. Tracing the evolution chiral symmetry, one finds that, in the region $\mu < 922$~MeV, the global minimum of $\Omega$ lies along the $\beta=0$ axis, as shown in Figs.~\ref{fig:with_x_muc_0} and~\ref{fig:with_x_muc_4}. For $922 < \mu < 1535$~MeV in set B, $\omega_{{ir}}$ gradually decreases as $\mu$ increases. When $\mu > 1535$~MeV, $\omega_{{ir}} = 0$, indicating that chiral symmetry is restored. 
In set A, when $\mu > 922$~MeV, $\omega_{ir}$ remains constantly equal to $0$, so the situation is similar to the case without $\Phi$, where the chiral symmetry remains permanently restored.

\section{Equation of state and neutron star structure}
\label{sec:neutron_stars}

We finally study the properties of neutron stars formed by baryonic matter described in the previous sections to see whether the present approach is reasonable. For this purpose, we first calculate the corresponding EOS. 

\begin{figure}[htb]
    \centering
    \begin{overpic}[width=0.8\linewidth]{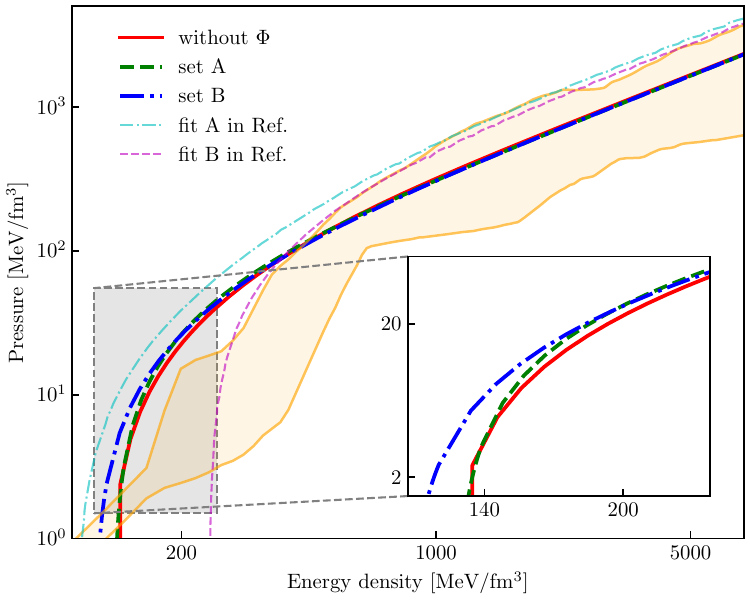}
        \put(20, 62.5){
            {\fontsize{6pt}{2pt}\selectfont \quad\quad\quad\quad\quad\hspace{-1.8mm}~\cite{Bartolini:2022rkl}}}
        \put(20, 58.5){
            {\fontsize{6pt}{2pt}\selectfont \quad\quad\quad\quad\quad\hspace{-1.8mm}~\cite{Bartolini:2022rkl}}}
        \end{overpic}
    \caption{ EOSs under three sets of parameters. 
    The orange region in the figure is the area constrained data~\cite{Bartolini:2023wis}. 
    %In 
    The results from different hard-wall models~\cite{Bartolini:2022rkl} are also shown for comparison. 
    }
    \label{fig:e_p}
\end{figure}

In the zero temperature limit, the pressure $P$ and energy density $\mathcal{E}$ 
can be expressed as: 

\be
& & PV =  - \Omega  = S = V \int dz \mathcal{L} \nonumber\\
& & {\cal E} = \frac{E}{V} =  - P + \mu \rho 
\label{eq:eos}
\ee
where $V=\int d^4 x$, $S$ is the action in Eqs.~\eqref{eq:action_ads} and~\eqref{eq:action_homo}. $\Omega$ is the grand potential. 

Our result of the EOS is plotted in Fig.~\ref{fig:e_p}. 
In this figure, the red-solid line shows EOS without scalar field $\Phi$, 
the green-dashed line shows EOS of set A, the blue-dot-dashed line shows EOS of set B. For comparison, we also plot the constraint from the present combined analysis~\cite{Bartolini:2023wis}. One can see that, the EOS obtained from the present hQCD model is align with the constraint at density $\gtrsim 3.0 \rho_0$ , but in the low-density region, it slightly exceeds the upper bound of the constrained region. This is understandable since in the present description of nucleon is soliton without spin which is valid in the large $N_c$ limit~\cite{Witten:1979kh}. A similar conclusion of the nuclear matter was found in the skyrmion crytal approach~\cite{Ma:2013ooa,Ma:2013ela,Shao:2022njr}. Comparing with the same hard-wall model approach without considering the pure gauge field $U(x)$~\cite{Bartolini:2022rkl}, one can see that the present approach leads to a significant improvement.

The neutron star properties can be obtained by solving the Tolman-Oppenheimer-Volkoff (TOV) equations
\be
& & \frac{{dP}}{{dr}} =  - G({\cal E} + P)\frac{{m + 4\pi {r^3}P}}{{r(r - 2Gm)}} \nonumber\\
& & \frac{{dm}}{{dr}} = 4\pi {r^2}{\cal E}
\label{eq:tov}
\ee
where $G$ is the gravitational constant, $m$ is mass density. According to these equations, given appropriate boundary conditions, 
the total mass $M$ and radius of a neutron star can be determined. 
Starting with a fixed central pressure $P(r=0)$ and central mass density 
$m(r=0)$ of the neutron star as initial conditions, the equations can 
be integrated until $P(R) = 0$ with $R$ being the radius of the neutron star. 
The total mass of the star with radius $R$ is $M=\int_0^R m dr$.

\cmb{}
{Since we regard baryons as state solitons in the work, the $P\geq 0$ region of the nuclear matter calculated above which contribute to the compact star starts from $\rho_0$. Therefore, in the calculation of the mass–radius relation of neutron stars, we match the above calculated EOS to the crust EOS from Refs.~\cite{Chabanat:1997qh, Chabanat:1997un, Douchin:2000kx, Baym:1971pw}. 
%All the results before this section are without the crust. 
The procedure of the matching is illustrated in App.~\ref{app:hybrid_eos}.
} For the TOV solver, we refer to the publicly available code from GitHub repository~\footnote{https://github.com/amotornenko/TOVsolver, https://github.com/jnoronhahostler/{Neutron\_Star\_EOS}}.
% (https://github.com/amotornenko/TOVsolver, https://github.com/jnoronhahostler/{Neutron\_Star\_EOS).

\begin{figure}[ht]
    \centering
    \includegraphics[width=0.8\linewidth]{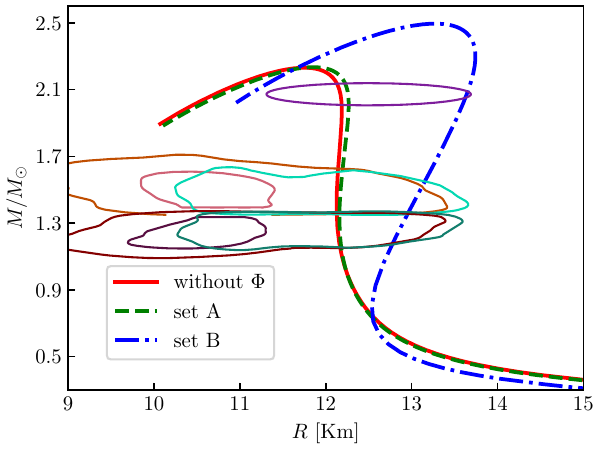}
    \caption{
        The mass-radius relation for three different parameter sets 
    and constraint data for neutron stars 
   ~\cite{Miller:2019cac, LIGOScientific:2020zkf, LIGOScientific:2018cki, Miller:2021qha, Riley:2021pdl}.
    }
    \label{fig:m_r}
\end{figure}

Following the above procedure, we obtain the mass-radius 
relation illustrated in Fig.~\ref{fig:m_r}. We also show 
some constraint data for neutron stars~\cite{Tan:2020ics, Miller:2019cac, LIGOScientific:2020zkf, LIGOScientific:2018cki, Miller:2021qha, Riley:2021pdl}. From this figure, we find that the mass-radius relation obtained from three sets of parameters show only slight difference if chiral symmetry restoration does not occur in the baryonic phase (the case in without $\Phi$ and set A), 
which is related to the similarity of the EOS in Fig.~\ref{fig:e_p}. 
This is because the fitted values of \( D \) are distributed within a very narrow range, causing the starting points of the equations of state (EOSs) to be very close to each other, shown in the lower-left corner of Fig.~\ref{fig:e_p}. However, at high densities, the behaviors of different EOSs converge into a similar region, shown in the upper-right corner of Fig.~\ref{fig:e_p}. The combination of nearby starting points and consistent high-density behavior results in only small differences among the EOSs under different parameters, leading to negligible differences in the resulting mass–radius relations. 
If chiral symmetry restoration occurs in the baryonic phase (the case in set B), the EOS and mass-radius relation will show relatively obvious changes. In this case, the scalar field leads to modifications of the D value (shown in the lower-left corner of Fig.~\ref{fig:e_p}). Within the low-density region, the EOS exhibits obvious distinctions from the previous case, which ultimately results in alterations to the mass-radius relations.

\section{summary and discussion}
\label{sec:summary}

In this work, we investigated the properties of nuclear matter 
and neutron stars formed by homogeneous nuclear matter using the hard-wall hQCD model including the CS terms. We explicitly illustrated the $\mu$-$\rho$ relation, the speed of sound, the EOS and mass-radius relation of neutron stars.

By using the homogeneous approximation including the holonomy that carries pion degrees of freedom, the chiral phase transition can be decoupled from the onset of baryons, which was not achieved in previous models. Using three different parameter sets, we found that, in the 
absence of chiral phase transition in baryonic phase, the results with and 
without scalar field $\Phi$ are similar. However, when a chiral 
phase transition occurs, the speed of sound and mass-radius 
relation changes 
significantly. All the mass-radius relation fall within 
the 90\% confidence 
interval of the constraint data, demonstrating the strong 
performance of our model. In addition, irrelevant to the existence of chiral phase transition, we also find that the speed of sound naturally converges to the conformal limit at the density relevant to cores of massive stars but the trace of energy-momentum tensor does not vanish which indicates the pseudoconformal structure and intrinsic manifestation of scale symmetry in compact star matter.

In this work, we determine $D = 1.41$ in the model without 
the scalar field by minimizing the grand potential $\Omega$ at $\mu = 922$~MeV. But in this case, $\rho_0 \approx 0.138$~fm$^{-3}$, deviates from the empirical value $0.155\pm 0.050$~fm$^{-3}$~\cite{Sedrakian:2022ata}. In future, 
we consider that breaking the $U(2)$ symmetry of gauge fields to
$SU(2) \times U(1)$ to achieve better  consistency with data.

At finite density, the definition of chiral condensate still lacks a clear analytical expression. Here, we use the value of $\omega_{ir}$ to represent the strength of the chiral condensate, which is merely a temporary compromise. It is interesting to attempt to derive the asymptotic behavior of the chiral condensate. In addition, we will also investigate the effect of explicit chiral symmetry breaking, which was not considered in this work on the chiral phase structure.

\cmb{}
{Finally, we want to say that, in this work, baryons were considered as static solitons without spin and isospin quantum numbers and therefore the dense matter is symmetric one. It is more realistic to study the isospin asymmetric matter by including the isospin chemical potential~\cite{Kovensky:2021kzl,Kovensky:2023mye,Bartolini:2023wis,Bartolini:2025sag} and spin of baryons~\cite{weigelChiralSolitonModels2008, mantonSkyrmionsTheoryNuclei2022}. This will be investigated in the future publication. 
%therefore we did not distinguish proton and neutron. Distinguishing between protons and neutrons requires introducing collective rotations of the soliton to quantize the system~\cite{weigelChiralSolitonModels2008, mantonSkyrmionsTheoryNuclei2022}, which is computationally complicated. Therefore, our results actually correspond to the case of isospin-symmetric nuclear matter. In future work, we will consider investigating the properties of isospin-asymmetric nuclear matter~\cite{Bartolini:2025sag} by other methods.
}

\section*{Acknowledgments}

The work of J. S. W. was supported by the National Natural Science Foundation of 
China (NSFC) under Grant No. 12347103 and the Fundamental Research 
Funds for the Central Universities.
The work of Y.~L. M. is supported in part by the National Science Foundation of China (NSFC) under Grant No. 12347103 and the National Key R\&D Program of China under Grant No. 2021YFC2202900.

\appendix

\section{Derivation of the equation of motion the pure gauge field}
\label{app}

\cmb{}
{
Following Refs.~\cite{Sakai:2004cn, PhysRevD.82.076010}, in the hard-wall model used here, HLS can be realized by choosing the gauge $L_z=R_z=0$~\cite{PhysRevD.89.115012}.
In this case, the UV boundary values of the gauge fields correspond to the Maurer--Cartan 1-forms:  
\begin{equation}
    l_\mu\big|_{z_{uv}}=i\xi_L\partial_\mu\xi^\dagger_L\,,\quad 
    r_\mu\big|_{z_{uv}}=i\xi_R\partial_\mu\xi^\dagger_R\,.
\end{equation}
The pion field $U(x)$ can be defined as~\cite{Harada:2003jx}: 
\begin{equation}
    U(x)=\xi^\dagger_L(x)\,\xi_R(x) = \exp(2i\pi(x)/f^2_\pi).
\end{equation}
%In the WSS model, one may adopt the gauge fixing $\xi^\dagger_L(x)=U(x)$ and $\xi_R(x)=1$, thereby reducing the HLS theory to the Skyrme model \cite{Sakai:2004cn}. 
After gauge fixing, one has
\begin{equation}
\label{gaugefixing}
    \xi^\dagger_L(x)=\xi_R(x)=\xi=\exp(i\pi(x)/f^2_\pi).  
\end{equation}
The pure gauge condition or the zero curvature condition yields
\begin{equation}
\label{eq:puregauge}
    \partial_\mu l_\nu-\partial_\nu l_\mu-i\left[l_\mu,l_\nu\right]=0,
    \qquad (l_\mu\leftrightarrow r_\mu)\,.
\end{equation}

%In order to obtain a homogeneous parametrization in Eqs.~(\ref{eq:homogeneous_01}- \ref{eq:homogeneous_04}), which is similar to that of Refs.~\cite{Bartolini:2022rkl,Fujii:2025umi}, we choose the gauge fixing Eq.~(\ref{gaugefixing}), which gives  
%\begin{equation}
%    L_\mu\big|_{z_{uv}}=i\xi^\dagger\partial_\mu \xi,\quad 
%    R_\mu\big|_{z_{uv}}=i \xi\partial_\mu \xi^\dagger.
%\end{equation}

Performing the Kaluza--Klein (KK) decomposition, integrating out the higher vector meson modes,
%appear~\cite{PhysRevD.89.115012}.  
%Neglecting the contributions from these vector mesons, 
the bulk gauge fields can be written as  
\begin{equation}
    L_\mu(x,z)=l_\mu(x)\varphi_L(z)\,,\quad 
    R_\mu(x,z)= r_\mu(x)\varphi_R(z)\,,
\end{equation}
where $l_\mu=i\xi^\dagger\partial_\mu\xi$ and $r_\mu=i\xi\partial_\mu\xi^\dagger$ satisfying Eq.~\eqref{eq:puregauge}
%are pure-gauge configurations satisfying  
%\begin{equation}\label{eq:puregauge}
%    \partial_\mu l_\nu-\partial_\nu l_\mu-i\left[l_\mu,l_\nu\right]=0,
%    \qquad (l_\mu\leftrightarrow r_\mu)\,, 
%\end{equation}
%in agreement with Eq.~(\ref{puregauge}).
Then, for a parity-even static solution, which is commonly considered in the literature~\cite{Domenech:2010aq,Pomarol:2008aa}, one has 
\begin{equation}
\hspace{-3mm}
    L_{i}(\bm{x},z)=l_i(\bm{x})\varphi(z)=-R_i(-\bm{x},z)=-r_i(-\bm{x})\varphi(z),
\end{equation}
which, together with Eq.~(\ref{eq:puregauge}), yields
\begin{align}
    L_{ij}(\bm{x},z)&=R_{ij}(-\bm{x},z) \nonumber\\
    & \hspace{-9mm} =\varphi(z)\Big(\partial_il_j(\bm{x})-\partial_jl_i(\bm{x})\Big)-i\varphi(z)^2\Big[l_i(\bm{x}),\,l_j(\bm{x})\Big]\nonumber\\
    & \hspace{-9mm} =i\Big[l_i(\bm{x}),l_j(\bm{x})\Big]\varphi(z)\Big(1-\varphi(z)\Big).
\end{align}
Then, by using Eq.~(\ref{eq:homogeneous_01}) one finally obtains Eq.~(\ref{eq:homogeneous_04}).

% end for cmb
}

\section{Construction of the hybrid EOS}
\label{app:hybrid_eos}

\begin{figure}[htbp]
    \centering
\includegraphics[width=0.8\linewidth]{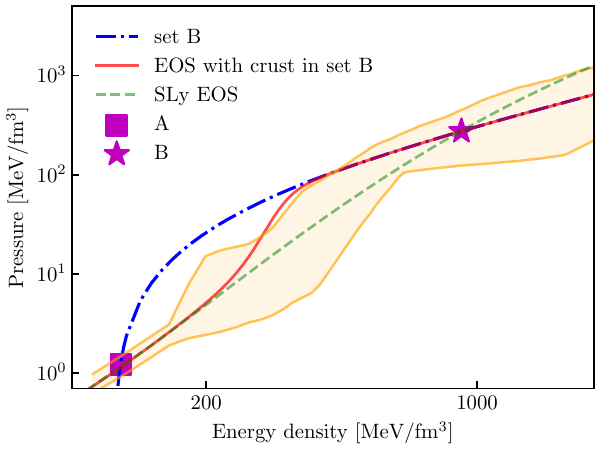}
    \caption{
    Construction of the hybrid EOS by using crossover transition. The hybrid EOS is in red-soild line, the SLy EOS is in green-dashed line, and the EOS in set B in blue-dot-dashed line.
    }
    \label{fig:eos_hybrid}
\end{figure}

\cmb{}
{
%Since we have attached an crust when calculating the mass–radius relation, the matching procedure is described as follows:

Our numerical calculation is based on a publicly available TOV equation code from  GitHub repository which uses the Gibbs construction~\cite{Bhattacharyya:2009fg} to connect the homogeneous EOS and crust. We modify the connection to a crossover scheme~\cite{Masuda:2012ed} in order to have smooth EOS.

Here, to illustrate the proceedure, we take the parameter set B as an example. In this case:
\begin{enumerate}
    \item The EOS of set B intersects with the SLy EOS at two points, denoted as points A and B in Fig.~\ref{fig:eos_hybrid}. 
    
    \item We adopt a crossover construction between the EOS of set B and the SLy EOS. %It is natural to choose points A and B as the starting and ending endpoints of the construction interval, 
    During the connection, we take into account two key considerations: (a) To use the EOS obtained from hQCD over as wide a density range as possible, and (b) To ensure that the resulting hybrid EOS remains within the constraint region as much as possible.

    \item We use the function \(\tanh(x)\) as the smoothing function for crossover construction, 
and $\bar{\cal{E}}=(\cal{E}_A+\cal{E}_B)/\text{2}$ as the central energy density of the transition region. $\cal{E}_A$ is the energy density at point A in Fig.~\ref{fig:eos_hybrid}, and $\cal{E}_B$ is that at point B. Then the pressure $P$ of the hybrid EOS is defined as:
\be
	\label{eq:f_for_crossover}
	f(\cal{E}) & = & \left(\tanh \Big( \dfrac{\cal{E}-\bar{\cal{E}}}{s} \Big) + 1 \right)/2, \nonumber\\
%	\label{eq:p_for_crossover}
	P & = & (1-f) \times P_{SLy} + f \times P_{setB}. 
\ee
In the Eq.~(\ref{eq:f_for_crossover}), $P_{SLy}$ and $P_{setB}$ are the pressure of EOS in SLy and set B at given energy density $\cal{E}$. $s$ is the scaling factor, which can be reasonably taken as $s=1$. 
\end{enumerate}
%We can adjust the parameter $s$ to make the EOS between points A and B recover to the crust and hQCD EOS more quickly or more slowly. As shown in Fig.~\ref{fig:eos_hybrid}, when \( s = 1 \), there is only a small transition region, which serves the purpose described in Step 2. Moreover, this construction also prevents the chiral restoration critical point in set B from being covered by the transition region.
The final hybrid EOS is shown as the red solid line in the Fig.~\ref{fig:eos_hybrid}.
%which makes use of all the hQCD equations of state within the constraint region and ensures that the majority of the hybrid EOS lies within the constraint region as much as possible.

% end for cmb
}

\bibliographystyle{apsrev4-2}
\bibliography{main}

%apsrev4-2.bst 2019-01-14 (MD) hand-edited version of apsrev4-1.bst
%Control: key (0)
%Control: author (72) initials jnrlst
%Control: editor formatted (1) identically to author
%Control: production of article title (-1) disabled
%Control: page (0) single
%Control: year (1) truncated
%Control: production of eprint (0) enabled
\begin{thebibliography}{83}%
\makeatletter
\providecommand \@ifxundefined [1]{%
 \@ifx{#1\undefined}
}%
\providecommand \@ifnum [1]{%
 \ifnum #1\expandafter \@firstoftwo
 \else \expandafter \@secondoftwo
 \fi
}%
\providecommand \@ifx [1]{%
 \ifx #1\expandafter \@firstoftwo
 \else \expandafter \@secondoftwo
 \fi
}%
\providecommand \natexlab [1]{#1}%
\providecommand \enquote  [1]{``#1''}%
\providecommand \bibnamefont  [1]{#1}%
\providecommand \bibfnamefont [1]{#1}%
\providecommand \citenamefont [1]{#1}%
\providecommand \href@noop [0]{\@secondoftwo}%
\providecommand \href [0]{\begingroup \@sanitize@url \@href}%
\providecommand \@href[1]{\@@startlink{#1}\@@href}%
\providecommand \@@href[1]{\endgroup#1\@@endlink}%
\providecommand \@sanitize@url [0]{\catcode `\\12\catcode `\$12\catcode
  `\&12\catcode `\#12\catcode `\^12\catcode `\_12\catcode `\%12\relax}%
\providecommand \@@startlink[1]{}%
\providecommand \@@endlink[0]{}%
\providecommand \url  [0]{\begingroup\@sanitize@url \@url }%
\providecommand \@url [1]{\endgroup\@href {#1}{\urlprefix }}%
\providecommand \urlprefix  [0]{URL }%
\providecommand \Eprint [0]{\href }%
\providecommand \doibase [0]{https://doi.org/}%
\providecommand \selectlanguage [0]{\@gobble}%
\providecommand \bibinfo  [0]{\@secondoftwo}%
\providecommand \bibfield  [0]{\@secondoftwo}%
\providecommand \translation [1]{[#1]}%
\providecommand \BibitemOpen [0]{}%
\providecommand \bibitemStop [0]{}%
\providecommand \bibitemNoStop [0]{.\EOS\space}%
\providecommand \EOS [0]{\spacefactor3000\relax}%
\providecommand \BibitemShut  [1]{\csname bibitem#1\endcsname}%
\let\auto@bib@innerbib\@empty
%</preamble>
\bibitem [{\citenamefont {Fukushima}\ and\ \citenamefont
  {Hatsuda}(2011)}]{Fukushima:2010bq}%
  \BibitemOpen
  \bibfield  {author} {\bibinfo {author} {\bibfnamefont {K.}~\bibnamefont
  {Fukushima}}\ and\ \bibinfo {author} {\bibfnamefont {T.}~\bibnamefont
  {Hatsuda}},\ }\href {https://doi.org/10.1088/0034-4885/74/1/014001}
  {\bibfield  {journal} {\bibinfo  {journal} {Rept. Prog. Phys.}\ }\textbf
  {\bibinfo {volume} {74}},\ \bibinfo {pages} {014001} (\bibinfo {year}
  {2011})},\ \Eprint {https://arxiv.org/abs/1005.4814} {arXiv:1005.4814
  [hep-ph]} \BibitemShut {NoStop}%
\bibitem [{\citenamefont {Lattimer}\ and\ \citenamefont
  {Prakash}(2016)}]{Lattimer:2015nhk}%
  \BibitemOpen
  \bibfield  {author} {\bibinfo {author} {\bibfnamefont {J.~M.}\ \bibnamefont
  {Lattimer}}\ and\ \bibinfo {author} {\bibfnamefont {M.}~\bibnamefont
  {Prakash}},\ }\href {https://doi.org/10.1016/j.physrep.2015.12.005}
  {\bibfield  {journal} {\bibinfo  {journal} {Phys. Rept.}\ }\textbf {\bibinfo
  {volume} {621}},\ \bibinfo {pages} {127} (\bibinfo {year} {2016})},\ \Eprint
  {https://arxiv.org/abs/1512.07820} {arXiv:1512.07820 [astro-ph.SR]}
  \BibitemShut {NoStop}%
\bibitem [{\citenamefont {Baym}\ \emph {et~al.}(2018)\citenamefont {Baym},
  \citenamefont {Hatsuda}, \citenamefont {Kojo}, \citenamefont {Powell},
  \citenamefont {Song},\ and\ \citenamefont {Takatsuka}}]{Baym:2017whm}%
  \BibitemOpen
  \bibfield  {author} {\bibinfo {author} {\bibfnamefont {G.}~\bibnamefont
  {Baym}}, \bibinfo {author} {\bibfnamefont {T.}~\bibnamefont {Hatsuda}},
  \bibinfo {author} {\bibfnamefont {T.}~\bibnamefont {Kojo}}, \bibinfo {author}
  {\bibfnamefont {P.~D.}\ \bibnamefont {Powell}}, \bibinfo {author}
  {\bibfnamefont {Y.}~\bibnamefont {Song}},\ and\ \bibinfo {author}
  {\bibfnamefont {T.}~\bibnamefont {Takatsuka}},\ }\href
  {https://doi.org/10.1088/1361-6633/aaae14} {\bibfield  {journal} {\bibinfo
  {journal} {Rept. Prog. Phys.}\ }\textbf {\bibinfo {volume} {81}},\ \bibinfo
  {pages} {056902} (\bibinfo {year} {2018})},\ \Eprint
  {https://arxiv.org/abs/1707.04966} {arXiv:1707.04966 [astro-ph.HE]}
  \BibitemShut {NoStop}%
\bibitem [{\citenamefont {Ma}\ and\ \citenamefont {Rho}(2020)}]{Ma:2019ery}%
  \BibitemOpen
  \bibfield  {author} {\bibinfo {author} {\bibfnamefont {Y.-L.}\ \bibnamefont
  {Ma}}\ and\ \bibinfo {author} {\bibfnamefont {M.}~\bibnamefont {Rho}},\
  }\href {https://doi.org/10.1016/j.ppnp.2020.103791} {\bibfield  {journal}
  {\bibinfo  {journal} {Prog. Part. Nucl. Phys.}\ }\textbf {\bibinfo {volume}
  {113}},\ \bibinfo {pages} {103791} (\bibinfo {year} {2020})},\ \Eprint
  {https://arxiv.org/abs/1909.05889} {arXiv:1909.05889 [nucl-th]} \BibitemShut
  {NoStop}%
\bibitem [{\citenamefont {Brandes}\ and\ \citenamefont
  {Weise}(2024)}]{Brandes:2023bob}%
  \BibitemOpen
  \bibfield  {author} {\bibinfo {author} {\bibfnamefont {L.}~\bibnamefont
  {Brandes}}\ and\ \bibinfo {author} {\bibfnamefont {W.}~\bibnamefont
  {Weise}},\ }\href {https://doi.org/10.3390/sym16010111} {\bibfield  {journal}
  {\bibinfo  {journal} {Symmetry}\ }\textbf {\bibinfo {volume} {16}},\ \bibinfo
  {pages} {111} (\bibinfo {year} {2024})},\ \Eprint
  {https://arxiv.org/abs/2312.11937} {arXiv:2312.11937 [nucl-th]} \BibitemShut
  {NoStop}%
\bibitem [{\citenamefont {Sorensen}\ \emph {et~al.}(2024)\citenamefont
  {Sorensen} \emph {et~al.}}]{Sorensen:2023zkk}%
  \BibitemOpen
  \bibfield  {author} {\bibinfo {author} {\bibfnamefont {A.}~\bibnamefont
  {Sorensen}} \emph {et~al.},\ }\href
  {https://doi.org/10.1016/j.ppnp.2023.104080} {\bibfield  {journal} {\bibinfo
  {journal} {Prog. Part. Nucl. Phys.}\ }\textbf {\bibinfo {volume} {134}},\
  \bibinfo {pages} {104080} (\bibinfo {year} {2024})},\ \Eprint
  {https://arxiv.org/abs/2301.13253} {arXiv:2301.13253 [nucl-th]} \BibitemShut
  {NoStop}%
\bibitem [{\citenamefont {Cai}\ and\ \citenamefont {Li}(2025)}]{Cai:2025nxn}%
  \BibitemOpen
  \bibfield  {author} {\bibinfo {author} {\bibfnamefont {B.-J.}\ \bibnamefont
  {Cai}}\ and\ \bibinfo {author} {\bibfnamefont {B.-A.}\ \bibnamefont {Li}},\
  }\href {https://doi.org/10.1140/epja/s10050-025-01507-7} {\bibfield
  {journal} {\bibinfo  {journal} {Eur. Phys. J. A}\ }\textbf {\bibinfo {volume}
  {61}},\ \bibinfo {pages} {55} (\bibinfo {year} {2025})},\ \Eprint
  {https://arxiv.org/abs/2501.18676} {arXiv:2501.18676 [astro-ph.HE]}
  \BibitemShut {NoStop}%
\bibitem [{\citenamefont {Maldacena}(1998)}]{Maldacena:1997re}%
  \BibitemOpen
  \bibfield  {author} {\bibinfo {author} {\bibfnamefont {J.~M.}\ \bibnamefont
  {Maldacena}},\ }\href {https://doi.org/10.4310/ATMP.1998.v2.n2.a1} {\bibfield
   {journal} {\bibinfo  {journal} {Adv. Theor. Math. Phys.}\ }\textbf {\bibinfo
  {volume} {2}},\ \bibinfo {pages} {231} (\bibinfo {year} {1998})},\ \Eprint
  {https://arxiv.org/abs/hep-th/9711200} {arXiv:hep-th/9711200} \BibitemShut
  {NoStop}%
\bibitem [{\citenamefont {Gubser}\ \emph {et~al.}(1998)\citenamefont {Gubser},
  \citenamefont {Klebanov},\ and\ \citenamefont {Polyakov}}]{Gubser:1998bc}%
  \BibitemOpen
  \bibfield  {author} {\bibinfo {author} {\bibfnamefont {S.~S.}\ \bibnamefont
  {Gubser}}, \bibinfo {author} {\bibfnamefont {I.~R.}\ \bibnamefont
  {Klebanov}},\ and\ \bibinfo {author} {\bibfnamefont {A.~M.}\ \bibnamefont
  {Polyakov}},\ }\href {https://doi.org/10.1016/S0370-2693(98)00377-3}
  {\bibfield  {journal} {\bibinfo  {journal} {Phys. Lett. B}\ }\textbf
  {\bibinfo {volume} {428}},\ \bibinfo {pages} {105} (\bibinfo {year}
  {1998})},\ \Eprint {https://arxiv.org/abs/hep-th/9802109}
  {arXiv:hep-th/9802109} \BibitemShut {NoStop}%
\bibitem [{\citenamefont {Witten}(1998)}]{Witten:1998qj}%
  \BibitemOpen
  \bibfield  {author} {\bibinfo {author} {\bibfnamefont {E.}~\bibnamefont
  {Witten}},\ }\href {https://doi.org/10.4310/ATMP.1998.v2.n2.a2} {\bibfield
  {journal} {\bibinfo  {journal} {Adv. Theor. Math. Phys.}\ }\textbf {\bibinfo
  {volume} {2}},\ \bibinfo {pages} {253} (\bibinfo {year} {1998})},\ \Eprint
  {https://arxiv.org/abs/hep-th/9802150} {arXiv:hep-th/9802150} \BibitemShut
  {NoStop}%
\bibitem [{\citenamefont {Natsuume}(2015)}]{Natsuume:2014sfa}%
  \BibitemOpen
  \bibfield  {author} {\bibinfo {author} {\bibfnamefont {M.}~\bibnamefont
  {Natsuume}},\ }\href {https://doi.org/10.1007/978-4-431-55441-7} {\emph
  {\bibinfo {title} {{AdS/CFT Duality User Guide}}}},\ Vol.\ \bibinfo {volume}
  {903}\ (\bibinfo {year} {2015})\ \Eprint {https://arxiv.org/abs/1409.3575}
  {arXiv:1409.3575 [hep-th]} \BibitemShut {NoStop}%
\bibitem [{\citenamefont {Ammon}\ and\ \citenamefont
  {Erdmenger}(2015)}]{Ammon:2015wua}%
  \BibitemOpen
  \bibfield  {author} {\bibinfo {author} {\bibfnamefont {M.}~\bibnamefont
  {Ammon}}\ and\ \bibinfo {author} {\bibfnamefont {J.}~\bibnamefont
  {Erdmenger}},\ }\href {https://doi.org/10.1017/CBO9780511846373} {\emph
  {\bibinfo {title} {{Gauge/gravity duality}: {Foundations and
  applications}}}}\ (\bibinfo  {publisher} {Cambridge University Press},\
  \bibinfo {address} {Cambridge},\ \bibinfo {year} {2015})\BibitemShut
  {NoStop}%
\bibitem [{\citenamefont {Kim}\ \emph {et~al.}(2013)\citenamefont {Kim},
  \citenamefont {Shin},\ and\ \citenamefont {Tsukioka}}]{Kim:2012ey}%
  \BibitemOpen
  \bibfield  {author} {\bibinfo {author} {\bibfnamefont {Y.}~\bibnamefont
  {Kim}}, \bibinfo {author} {\bibfnamefont {I.~J.}\ \bibnamefont {Shin}},\ and\
  \bibinfo {author} {\bibfnamefont {T.}~\bibnamefont {Tsukioka}},\ }\href
  {https://doi.org/10.1016/j.ppnp.2012.09.002} {\bibfield  {journal} {\bibinfo
  {journal} {Prog. Part. Nucl. Phys.}\ }\textbf {\bibinfo {volume} {68}},\
  \bibinfo {pages} {55} (\bibinfo {year} {2013})},\ \Eprint
  {https://arxiv.org/abs/1205.4852} {arXiv:1205.4852 [hep-ph]} \BibitemShut
  {NoStop}%
\bibitem [{\citenamefont {Sakai}\ and\ \citenamefont
  {Sugimoto}(2005{\natexlab{a}})}]{Sakai:2004cn}%
  \BibitemOpen
  \bibfield  {author} {\bibinfo {author} {\bibfnamefont {T.}~\bibnamefont
  {Sakai}}\ and\ \bibinfo {author} {\bibfnamefont {S.}~\bibnamefont
  {Sugimoto}},\ }\href {https://doi.org/10.1143/PTP.113.843} {\bibfield
  {journal} {\bibinfo  {journal} {Prog. Theor. Phys.}\ }\textbf {\bibinfo
  {volume} {113}},\ \bibinfo {pages} {843} (\bibinfo {year}
  {2005}{\natexlab{a}})},\ \Eprint {https://arxiv.org/abs/hep-th/0412141}
  {arXiv:hep-th/0412141} \BibitemShut {NoStop}%
\bibitem [{\citenamefont {Sakai}\ and\ \citenamefont
  {Sugimoto}(2005{\natexlab{b}})}]{Sakai:2005yt}%
  \BibitemOpen
  \bibfield  {author} {\bibinfo {author} {\bibfnamefont {T.}~\bibnamefont
  {Sakai}}\ and\ \bibinfo {author} {\bibfnamefont {S.}~\bibnamefont
  {Sugimoto}},\ }\href {https://doi.org/10.1143/PTP.114.1083} {\bibfield
  {journal} {\bibinfo  {journal} {Prog. Theor. Phys.}\ }\textbf {\bibinfo
  {volume} {114}},\ \bibinfo {pages} {1083} (\bibinfo {year}
  {2005}{\natexlab{b}})},\ \Eprint {https://arxiv.org/abs/hep-th/0507073}
  {arXiv:hep-th/0507073} \BibitemShut {NoStop}%
\bibitem [{\citenamefont {Erlich}\ \emph {et~al.}(2005)\citenamefont {Erlich},
  \citenamefont {Katz}, \citenamefont {Son},\ and\ \citenamefont
  {Stephanov}}]{Erlich:2005qh}%
  \BibitemOpen
  \bibfield  {author} {\bibinfo {author} {\bibfnamefont {J.}~\bibnamefont
  {Erlich}}, \bibinfo {author} {\bibfnamefont {E.}~\bibnamefont {Katz}},
  \bibinfo {author} {\bibfnamefont {D.~T.}\ \bibnamefont {Son}},\ and\ \bibinfo
  {author} {\bibfnamefont {M.~A.}\ \bibnamefont {Stephanov}},\ }\href
  {https://doi.org/10.1103/PhysRevLett.95.261602} {\bibfield  {journal}
  {\bibinfo  {journal} {Phys. Rev. Lett.}\ }\textbf {\bibinfo {volume} {95}},\
  \bibinfo {pages} {261602} (\bibinfo {year} {2005})},\ \Eprint
  {https://arxiv.org/abs/hep-ph/0501128} {arXiv:hep-ph/0501128} \BibitemShut
  {NoStop}%
\bibitem [{\citenamefont {Da~Rold}\ and\ \citenamefont
  {Pomarol}(2005)}]{DaRold:2005mxj}%
  \BibitemOpen
  \bibfield  {author} {\bibinfo {author} {\bibfnamefont {L.}~\bibnamefont
  {Da~Rold}}\ and\ \bibinfo {author} {\bibfnamefont {A.}~\bibnamefont
  {Pomarol}},\ }\href {https://doi.org/10.1016/j.nuclphysb.2005.05.009}
  {\bibfield  {journal} {\bibinfo  {journal} {Nucl. Phys. B}\ }\textbf
  {\bibinfo {volume} {721}},\ \bibinfo {pages} {79} (\bibinfo {year} {2005})},\
  \Eprint {https://arxiv.org/abs/hep-ph/0501218} {arXiv:hep-ph/0501218}
  \BibitemShut {NoStop}%
\bibitem [{\citenamefont {Hoyos}\ \emph {et~al.}(2022)\citenamefont {Hoyos},
  \citenamefont {Jokela},\ and\ \citenamefont {Vuorinen}}]{Hoyos:2021uff}%
  \BibitemOpen
  \bibfield  {author} {\bibinfo {author} {\bibfnamefont {C.}~\bibnamefont
  {Hoyos}}, \bibinfo {author} {\bibfnamefont {N.}~\bibnamefont {Jokela}},\ and\
  \bibinfo {author} {\bibfnamefont {A.}~\bibnamefont {Vuorinen}},\ }\href
  {https://doi.org/10.1016/j.ppnp.2022.103972} {\bibfield  {journal} {\bibinfo
  {journal} {Prog. Part. Nucl. Phys.}\ }\textbf {\bibinfo {volume} {126}},\
  \bibinfo {pages} {103972} (\bibinfo {year} {2022})},\ \Eprint
  {https://arxiv.org/abs/2112.08422} {arXiv:2112.08422 [hep-th]} \BibitemShut
  {NoStop}%
\bibitem [{\citenamefont {Rougemont}\ \emph {et~al.}(2024)\citenamefont
  {Rougemont}, \citenamefont {Grefa}, \citenamefont {Hippert}, \citenamefont
  {Noronha}, \citenamefont {Noronha-Hostler}, \citenamefont {Portillo},\ and\
  \citenamefont {Ratti}}]{Rougemont:2023gfz}%
  \BibitemOpen
  \bibfield  {author} {\bibinfo {author} {\bibfnamefont {R.}~\bibnamefont
  {Rougemont}}, \bibinfo {author} {\bibfnamefont {J.}~\bibnamefont {Grefa}},
  \bibinfo {author} {\bibfnamefont {M.}~\bibnamefont {Hippert}}, \bibinfo
  {author} {\bibfnamefont {J.}~\bibnamefont {Noronha}}, \bibinfo {author}
  {\bibfnamefont {J.}~\bibnamefont {Noronha-Hostler}}, \bibinfo {author}
  {\bibfnamefont {I.}~\bibnamefont {Portillo}},\ and\ \bibinfo {author}
  {\bibfnamefont {C.}~\bibnamefont {Ratti}},\ }\href
  {https://doi.org/10.1016/j.ppnp.2023.104093} {\bibfield  {journal} {\bibinfo
  {journal} {Prog. Part. Nucl. Phys.}\ }\textbf {\bibinfo {volume} {135}},\
  \bibinfo {pages} {104093} (\bibinfo {year} {2024})},\ \Eprint
  {https://arxiv.org/abs/2307.03885} {arXiv:2307.03885 [nucl-th]} \BibitemShut
  {NoStop}%
\bibitem [{\citenamefont {Jarvinen}(2023)}]{Jarvinen:2023jbr}%
  \BibitemOpen
  \bibfield  {author} {\bibinfo {author} {\bibfnamefont {M.}~\bibnamefont
  {Jarvinen}},\ }\href {https://doi.org/10.22128/jhap.2023.695.1054} {\bibfield
   {journal} {\bibinfo  {journal} {JHAP}\ }\textbf {\bibinfo {volume} {3}},\
  \bibinfo {pages} {1} (\bibinfo {year} {2023})},\ \Eprint
  {https://arxiv.org/abs/2307.01745} {arXiv:2307.01745 [hep-ph]} \BibitemShut
  {NoStop}%
\bibitem [{\citenamefont {Gherghetta}\ \emph {et~al.}(2009)\citenamefont
  {Gherghetta}, \citenamefont {Kapusta},\ and\ \citenamefont
  {Kelley}}]{Gherghetta:2009ac}%
  \BibitemOpen
  \bibfield  {author} {\bibinfo {author} {\bibfnamefont {T.}~\bibnamefont
  {Gherghetta}}, \bibinfo {author} {\bibfnamefont {J.~I.}\ \bibnamefont
  {Kapusta}},\ and\ \bibinfo {author} {\bibfnamefont {T.~M.}\ \bibnamefont
  {Kelley}},\ }\href {https://doi.org/10.1103/PhysRevD.79.076003} {\bibfield
  {journal} {\bibinfo  {journal} {Phys. Rev. D}\ }\textbf {\bibinfo {volume}
  {79}},\ \bibinfo {pages} {076003} (\bibinfo {year} {2009})},\ \Eprint
  {https://arxiv.org/abs/0902.1998} {arXiv:0902.1998 [hep-ph]} \BibitemShut
  {NoStop}%
\bibitem [{\citenamefont {Chelabi}\ \emph {et~al.}(2016)\citenamefont
  {Chelabi}, \citenamefont {Fang}, \citenamefont {Huang}, \citenamefont {Li},\
  and\ \citenamefont {Wu}}]{Chelabi:2015gpc}%
  \BibitemOpen
  \bibfield  {author} {\bibinfo {author} {\bibfnamefont {K.}~\bibnamefont
  {Chelabi}}, \bibinfo {author} {\bibfnamefont {Z.}~\bibnamefont {Fang}},
  \bibinfo {author} {\bibfnamefont {M.}~\bibnamefont {Huang}}, \bibinfo
  {author} {\bibfnamefont {D.}~\bibnamefont {Li}},\ and\ \bibinfo {author}
  {\bibfnamefont {Y.-L.}\ \bibnamefont {Wu}},\ }\href
  {https://doi.org/10.1007/JHEP04(2016)036} {\bibfield  {journal} {\bibinfo
  {journal} {JHEP}\ }\textbf {\bibinfo {volume} {04}},\ \bibinfo {pages}
  {036}},\ \Eprint {https://arxiv.org/abs/1512.06493} {arXiv:1512.06493
  [hep-ph]} \BibitemShut {NoStop}%
\bibitem [{\citenamefont {Ghoroku}\ and\ \citenamefont
  {Yahiro}(2004)}]{Ghoroku:2004sp}%
  \BibitemOpen
  \bibfield  {author} {\bibinfo {author} {\bibfnamefont {K.}~\bibnamefont
  {Ghoroku}}\ and\ \bibinfo {author} {\bibfnamefont {M.}~\bibnamefont
  {Yahiro}},\ }\href {https://doi.org/10.1016/j.physletb.2004.10.048}
  {\bibfield  {journal} {\bibinfo  {journal} {Phys. Lett. B}\ }\textbf
  {\bibinfo {volume} {604}},\ \bibinfo {pages} {235} (\bibinfo {year}
  {2004})},\ \Eprint {https://arxiv.org/abs/hep-th/0408040}
  {arXiv:hep-th/0408040} \BibitemShut {NoStop}%
\bibitem [{\citenamefont {Cherman}\ \emph {et~al.}(2009)\citenamefont
  {Cherman}, \citenamefont {Cohen},\ and\ \citenamefont
  {Werbos}}]{Cherman:2008eh}%
  \BibitemOpen
  \bibfield  {author} {\bibinfo {author} {\bibfnamefont {A.}~\bibnamefont
  {Cherman}}, \bibinfo {author} {\bibfnamefont {T.~D.}\ \bibnamefont {Cohen}},\
  and\ \bibinfo {author} {\bibfnamefont {E.~S.}\ \bibnamefont {Werbos}},\
  }\href {https://doi.org/10.1103/PhysRevC.79.045203} {\bibfield  {journal}
  {\bibinfo  {journal} {Phys. Rev. C}\ }\textbf {\bibinfo {volume} {79}},\
  \bibinfo {pages} {045203} (\bibinfo {year} {2009})},\ \Eprint
  {https://arxiv.org/abs/0804.1096} {arXiv:0804.1096 [hep-ph]} \BibitemShut
  {NoStop}%
\bibitem [{\citenamefont {Kim}\ \emph {et~al.}(2010)\citenamefont {Kim},
  \citenamefont {Kim},\ and\ \citenamefont {Yakhshiev}}]{Kim:2009qs}%
  \BibitemOpen
  \bibfield  {author} {\bibinfo {author} {\bibfnamefont {H.-C.}\ \bibnamefont
  {Kim}}, \bibinfo {author} {\bibfnamefont {Y.}~\bibnamefont {Kim}},\ and\
  \bibinfo {author} {\bibfnamefont {U.~T.}\ \bibnamefont {Yakhshiev}},\ }\href
  {https://doi.org/10.1088/1674-1137/34/9/089} {\bibfield  {journal} {\bibinfo
  {journal} {Chin. Phys. C}\ }\textbf {\bibinfo {volume} {34}},\ \bibinfo
  {pages} {1520} (\bibinfo {year} {2010})},\ \Eprint
  {https://arxiv.org/abs/0912.1202} {arXiv:0912.1202 [hep-ph]} \BibitemShut
  {NoStop}%
\bibitem [{\citenamefont {Karch}\ \emph {et~al.}(2006)\citenamefont {Karch},
  \citenamefont {Katz}, \citenamefont {Son},\ and\ \citenamefont
  {Stephanov}}]{Karch:2006pv}%
  \BibitemOpen
  \bibfield  {author} {\bibinfo {author} {\bibfnamefont {A.}~\bibnamefont
  {Karch}}, \bibinfo {author} {\bibfnamefont {E.}~\bibnamefont {Katz}},
  \bibinfo {author} {\bibfnamefont {D.~T.}\ \bibnamefont {Son}},\ and\ \bibinfo
  {author} {\bibfnamefont {M.~A.}\ \bibnamefont {Stephanov}},\ }\href
  {https://doi.org/10.1103/PhysRevD.74.015005} {\bibfield  {journal} {\bibinfo
  {journal} {Phys. Rev. D}\ }\textbf {\bibinfo {volume} {74}},\ \bibinfo
  {pages} {015005} (\bibinfo {year} {2006})},\ \Eprint
  {https://arxiv.org/abs/hep-ph/0602229} {arXiv:hep-ph/0602229} \BibitemShut
  {NoStop}%
\bibitem [{\citenamefont {Park}\ \emph {et~al.}(2011)\citenamefont {Park},
  \citenamefont {Gwak}, \citenamefont {Lee}, \citenamefont {Ko},\ and\
  \citenamefont {Shin}}]{Park:2011qq}%
  \BibitemOpen
  \bibfield  {author} {\bibinfo {author} {\bibfnamefont {C.}~\bibnamefont
  {Park}}, \bibinfo {author} {\bibfnamefont {D.-Y.}\ \bibnamefont {Gwak}},
  \bibinfo {author} {\bibfnamefont {B.-H.}\ \bibnamefont {Lee}}, \bibinfo
  {author} {\bibfnamefont {Y.}~\bibnamefont {Ko}},\ and\ \bibinfo {author}
  {\bibfnamefont {S.}~\bibnamefont {Shin}},\ }\href
  {https://doi.org/10.1103/PhysRevD.84.046007} {\bibfield  {journal} {\bibinfo
  {journal} {Phys. Rev. D}\ }\textbf {\bibinfo {volume} {84}},\ \bibinfo
  {pages} {046007} (\bibinfo {year} {2011})},\ \bibinfo {note} {[Erratum:
  Phys.Rev.D 90, 129902 (2014)]},\ \Eprint {https://arxiv.org/abs/1104.4182}
  {arXiv:1104.4182 [hep-th]} \BibitemShut {NoStop}%
\bibitem [{\citenamefont {Colangelo}\ \emph
  {et~al.}(2012{\natexlab{a}})\citenamefont {Colangelo}, \citenamefont
  {Giannuzzi}, \citenamefont {Nicotri},\ and\ \citenamefont
  {Tangorra}}]{Colangelo:2011sr}%
  \BibitemOpen
  \bibfield  {author} {\bibinfo {author} {\bibfnamefont {P.}~\bibnamefont
  {Colangelo}}, \bibinfo {author} {\bibfnamefont {F.}~\bibnamefont
  {Giannuzzi}}, \bibinfo {author} {\bibfnamefont {S.}~\bibnamefont {Nicotri}},\
  and\ \bibinfo {author} {\bibfnamefont {V.}~\bibnamefont {Tangorra}},\ }\href
  {https://doi.org/10.1140/epjc/s10052-012-2096-9} {\bibfield  {journal}
  {\bibinfo  {journal} {Eur. Phys. J. C}\ }\textbf {\bibinfo {volume} {72}},\
  \bibinfo {pages} {2096} (\bibinfo {year} {2012}{\natexlab{a}})},\ \Eprint
  {https://arxiv.org/abs/1112.4402} {arXiv:1112.4402 [hep-ph]} \BibitemShut
  {NoStop}%
\bibitem [{\citenamefont {Colangelo}\ \emph
  {et~al.}(2012{\natexlab{b}})\citenamefont {Colangelo}, \citenamefont
  {Giannuzzi},\ and\ \citenamefont {Nicotri}}]{Colangelo:2012jy}%
  \BibitemOpen
  \bibfield  {author} {\bibinfo {author} {\bibfnamefont {P.}~\bibnamefont
  {Colangelo}}, \bibinfo {author} {\bibfnamefont {F.}~\bibnamefont
  {Giannuzzi}},\ and\ \bibinfo {author} {\bibfnamefont {S.}~\bibnamefont
  {Nicotri}},\ }\href {https://doi.org/10.1007/JHEP05(2012)076} {\bibfield
  {journal} {\bibinfo  {journal} {JHEP}\ }\textbf {\bibinfo {volume} {05}},\
  \bibinfo {pages} {076}},\ \Eprint {https://arxiv.org/abs/1201.1564}
  {arXiv:1201.1564 [hep-ph]} \BibitemShut {NoStop}%
\bibitem [{\citenamefont {Bartolini}\ \emph {et~al.}(2022)\citenamefont
  {Bartolini}, \citenamefont {Gudnason}, \citenamefont {Leutgeb},\ and\
  \citenamefont {Rebhan}}]{Bartolini:2022rkl}%
  \BibitemOpen
  \bibfield  {author} {\bibinfo {author} {\bibfnamefont {L.}~\bibnamefont
  {Bartolini}}, \bibinfo {author} {\bibfnamefont {S.~B.}\ \bibnamefont
  {Gudnason}}, \bibinfo {author} {\bibfnamefont {J.}~\bibnamefont {Leutgeb}},\
  and\ \bibinfo {author} {\bibfnamefont {A.}~\bibnamefont {Rebhan}},\ }\href
  {https://doi.org/10.1103/PhysRevD.105.126014} {\bibfield  {journal} {\bibinfo
   {journal} {Phys. Rev. D}\ }\textbf {\bibinfo {volume} {105}},\ \bibinfo
  {pages} {126014} (\bibinfo {year} {2022})},\ \Eprint
  {https://arxiv.org/abs/2202.12845} {arXiv:2202.12845 [hep-th]} \BibitemShut
  {NoStop}%
\bibitem [{\citenamefont {Kovensky}\ \emph {et~al.}(2022)\citenamefont
  {Kovensky}, \citenamefont {Poole},\ and\ \citenamefont
  {Schmitt}}]{Kovensky:2021kzl}%
  \BibitemOpen
  \bibfield  {author} {\bibinfo {author} {\bibfnamefont {N.}~\bibnamefont
  {Kovensky}}, \bibinfo {author} {\bibfnamefont {A.}~\bibnamefont {Poole}},\
  and\ \bibinfo {author} {\bibfnamefont {A.}~\bibnamefont {Schmitt}},\ }\href
  {https://doi.org/10.1103/PhysRevD.105.034022} {\bibfield  {journal} {\bibinfo
   {journal} {Phys. Rev. D}\ }\textbf {\bibinfo {volume} {105}},\ \bibinfo
  {pages} {034022} (\bibinfo {year} {2022})},\ \Eprint
  {https://arxiv.org/abs/2111.03374} {arXiv:2111.03374 [hep-ph]} \BibitemShut
  {NoStop}%
\bibitem [{\citenamefont {Jokela}\ \emph {et~al.}(2019)\citenamefont {Jokela},
  \citenamefont {J\"arvinen},\ and\ \citenamefont {Remes}}]{Jokela:2018ers}%
  \BibitemOpen
  \bibfield  {author} {\bibinfo {author} {\bibfnamefont {N.}~\bibnamefont
  {Jokela}}, \bibinfo {author} {\bibfnamefont {M.}~\bibnamefont {J\"arvinen}},\
  and\ \bibinfo {author} {\bibfnamefont {J.}~\bibnamefont {Remes}},\ }\href
  {https://doi.org/10.1007/JHEP03(2019)041} {\bibfield  {journal} {\bibinfo
  {journal} {JHEP}\ }\textbf {\bibinfo {volume} {03}},\ \bibinfo {pages}
  {041}},\ \Eprint {https://arxiv.org/abs/1809.07770} {arXiv:1809.07770
  [hep-ph]} \BibitemShut {NoStop}%
\bibitem [{\citenamefont {Elliot-Ripley}\ \emph {et~al.}(2016)\citenamefont
  {Elliot-Ripley}, \citenamefont {Sutcliffe},\ and\ \citenamefont
  {Zamaklar}}]{Elliot-Ripley:2016uwb}%
  \BibitemOpen
  \bibfield  {author} {\bibinfo {author} {\bibfnamefont {M.}~\bibnamefont
  {Elliot-Ripley}}, \bibinfo {author} {\bibfnamefont {P.}~\bibnamefont
  {Sutcliffe}},\ and\ \bibinfo {author} {\bibfnamefont {M.}~\bibnamefont
  {Zamaklar}},\ }\href {https://doi.org/10.1007/JHEP10(2016)088} {\bibfield
  {journal} {\bibinfo  {journal} {JHEP}\ }\textbf {\bibinfo {volume} {10}},\
  \bibinfo {pages} {088}},\ \Eprint {https://arxiv.org/abs/1607.04832}
  {arXiv:1607.04832 [hep-th]} \BibitemShut {NoStop}%
\bibitem [{\citenamefont {Randall}\ and\ \citenamefont
  {Sundrum}(1999)}]{Randall:1999ee}%
  \BibitemOpen
  \bibfield  {author} {\bibinfo {author} {\bibfnamefont {L.}~\bibnamefont
  {Randall}}\ and\ \bibinfo {author} {\bibfnamefont {R.}~\bibnamefont
  {Sundrum}},\ }\href {https://doi.org/10.1103/PhysRevLett.83.3370} {\bibfield
  {journal} {\bibinfo  {journal} {Phys. Rev. Lett.}\ }\textbf {\bibinfo
  {volume} {83}},\ \bibinfo {pages} {3370} (\bibinfo {year} {1999})},\ \Eprint
  {https://arxiv.org/abs/hep-ph/9905221} {arXiv:hep-ph/9905221} \BibitemShut
  {NoStop}%
\bibitem [{\citenamefont {Domenech}\ \emph {et~al.}(2011)\citenamefont
  {Domenech}, \citenamefont {Panico},\ and\ \citenamefont
  {Wulzer}}]{Domenech:2010aq}%
  \BibitemOpen
  \bibfield  {author} {\bibinfo {author} {\bibfnamefont {O.}~\bibnamefont
  {Domenech}}, \bibinfo {author} {\bibfnamefont {G.}~\bibnamefont {Panico}},\
  and\ \bibinfo {author} {\bibfnamefont {A.}~\bibnamefont {Wulzer}},\ }\href
  {https://doi.org/10.1016/j.nuclphysa.2011.02.002} {\bibfield  {journal}
  {\bibinfo  {journal} {Nucl. Phys. A}\ }\textbf {\bibinfo {volume} {853}},\
  \bibinfo {pages} {97} (\bibinfo {year} {2011})},\ \Eprint
  {https://arxiv.org/abs/1009.0711} {arXiv:1009.0711 [hep-ph]} \BibitemShut
  {NoStop}%
\bibitem [{\citenamefont {Da~Rold}\ and\ \citenamefont
  {Pomarol}(2006)}]{DaRold:2005vr}%
  \BibitemOpen
  \bibfield  {author} {\bibinfo {author} {\bibfnamefont {L.}~\bibnamefont
  {Da~Rold}}\ and\ \bibinfo {author} {\bibfnamefont {A.}~\bibnamefont
  {Pomarol}},\ }\href {https://doi.org/10.1088/1126-6708/2006/01/157}
  {\bibfield  {journal} {\bibinfo  {journal} {JHEP}\ }\textbf {\bibinfo
  {volume} {01}},\ \bibinfo {pages} {157}},\ \Eprint
  {https://arxiv.org/abs/hep-ph/0510268} {arXiv:hep-ph/0510268} \BibitemShut
  {NoStop}%
\bibitem [{\citenamefont {Fang}\ \emph {et~al.}(2016)\citenamefont {Fang},
  \citenamefont {Wu},\ and\ \citenamefont {Zhang}}]{Fang:2016nfj}%
  \BibitemOpen
  \bibfield  {author} {\bibinfo {author} {\bibfnamefont {Z.}~\bibnamefont
  {Fang}}, \bibinfo {author} {\bibfnamefont {Y.-L.}\ \bibnamefont {Wu}},\ and\
  \bibinfo {author} {\bibfnamefont {L.}~\bibnamefont {Zhang}},\ }\href
  {https://doi.org/10.1016/j.physletb.2016.09.009} {\bibfield  {journal}
  {\bibinfo  {journal} {Phys. Lett. B}\ }\textbf {\bibinfo {volume} {762}},\
  \bibinfo {pages} {86} (\bibinfo {year} {2016})},\ \Eprint
  {https://arxiv.org/abs/1604.02571} {arXiv:1604.02571 [hep-ph]} \BibitemShut
  {NoStop}%
\bibitem [{\citenamefont {Fang}\ \emph {et~al.}(2019)\citenamefont {Fang},
  \citenamefont {Wu},\ and\ \citenamefont {Zhang}}]{Fang:2018axm}%
  \BibitemOpen
  \bibfield  {author} {\bibinfo {author} {\bibfnamefont {Z.}~\bibnamefont
  {Fang}}, \bibinfo {author} {\bibfnamefont {Y.-L.}\ \bibnamefont {Wu}},\ and\
  \bibinfo {author} {\bibfnamefont {L.}~\bibnamefont {Zhang}},\ }\href
  {https://doi.org/10.1103/PhysRevD.99.034028} {\bibfield  {journal} {\bibinfo
  {journal} {Phys. Rev. D}\ }\textbf {\bibinfo {volume} {99}},\ \bibinfo
  {pages} {034028} (\bibinfo {year} {2019})},\ \Eprint
  {https://arxiv.org/abs/1810.12525} {arXiv:1810.12525 [hep-ph]} \BibitemShut
  {NoStop}%
\bibitem [{\citenamefont {Evans}\ and\ \citenamefont
  {Shock}(2004)}]{Evans:2004ia}%
  \BibitemOpen
  \bibfield  {author} {\bibinfo {author} {\bibfnamefont {N.~J.}\ \bibnamefont
  {Evans}}\ and\ \bibinfo {author} {\bibfnamefont {J.~P.}\ \bibnamefont
  {Shock}},\ }\href {https://doi.org/10.1103/PhysRevD.70.046002} {\bibfield
  {journal} {\bibinfo  {journal} {Phys. Rev. D}\ }\textbf {\bibinfo {volume}
  {70}},\ \bibinfo {pages} {046002} (\bibinfo {year} {2004})},\ \Eprint
  {https://arxiv.org/abs/hep-th/0403279} {arXiv:hep-th/0403279} \BibitemShut
  {NoStop}%
\bibitem [{\citenamefont {Kim}\ \emph {et~al.}(2008{\natexlab{a}})\citenamefont
  {Kim}, \citenamefont {Lee},\ and\ \citenamefont {Yee}}]{Kim:2007xi}%
  \BibitemOpen
  \bibfield  {author} {\bibinfo {author} {\bibfnamefont {Y.}~\bibnamefont
  {Kim}}, \bibinfo {author} {\bibfnamefont {C.-H.}\ \bibnamefont {Lee}},\ and\
  \bibinfo {author} {\bibfnamefont {H.-U.}\ \bibnamefont {Yee}},\ }\href
  {https://doi.org/10.1103/PhysRevD.77.085030} {\bibfield  {journal} {\bibinfo
  {journal} {Phys. Rev. D}\ }\textbf {\bibinfo {volume} {77}},\ \bibinfo
  {pages} {085030} (\bibinfo {year} {2008}{\natexlab{a}})},\ \Eprint
  {https://arxiv.org/abs/0707.2637} {arXiv:0707.2637 [hep-ph]} \BibitemShut
  {NoStop}%
\bibitem [{\citenamefont {Ghoroku}\ \emph {et~al.}(2013)\citenamefont
  {Ghoroku}, \citenamefont {Kubo}, \citenamefont {Tachibana}, \citenamefont
  {Taminato},\ and\ \citenamefont {Toyoda}}]{Ghoroku:2012am}%
  \BibitemOpen
  \bibfield  {author} {\bibinfo {author} {\bibfnamefont {K.}~\bibnamefont
  {Ghoroku}}, \bibinfo {author} {\bibfnamefont {K.}~\bibnamefont {Kubo}},
  \bibinfo {author} {\bibfnamefont {M.}~\bibnamefont {Tachibana}}, \bibinfo
  {author} {\bibfnamefont {T.}~\bibnamefont {Taminato}},\ and\ \bibinfo
  {author} {\bibfnamefont {F.}~\bibnamefont {Toyoda}},\ }\href
  {https://doi.org/10.1103/PhysRevD.87.066006} {\bibfield  {journal} {\bibinfo
  {journal} {Phys. Rev. D}\ }\textbf {\bibinfo {volume} {87}},\ \bibinfo
  {pages} {066006} (\bibinfo {year} {2013})},\ \Eprint
  {https://arxiv.org/abs/1211.2499} {arXiv:1211.2499 [hep-th]} \BibitemShut
  {NoStop}%
\bibitem [{\citenamefont {Ghoroku}\ \emph {et~al.}(2014)\citenamefont
  {Ghoroku}, \citenamefont {Kubo}, \citenamefont {Tachibana},\ and\
  \citenamefont {Toyoda}}]{Ghoroku:2013gja}%
  \BibitemOpen
  \bibfield  {author} {\bibinfo {author} {\bibfnamefont {K.}~\bibnamefont
  {Ghoroku}}, \bibinfo {author} {\bibfnamefont {K.}~\bibnamefont {Kubo}},
  \bibinfo {author} {\bibfnamefont {M.}~\bibnamefont {Tachibana}},\ and\
  \bibinfo {author} {\bibfnamefont {F.}~\bibnamefont {Toyoda}},\ }\href
  {https://doi.org/10.1142/S0217751X14500602} {\bibfield  {journal} {\bibinfo
  {journal} {Int. J. Mod. Phys. A}\ }\textbf {\bibinfo {volume} {29}},\
  \bibinfo {pages} {1450060} (\bibinfo {year} {2014})},\ \Eprint
  {https://arxiv.org/abs/1311.1598} {arXiv:1311.1598 [hep-th]} \BibitemShut
  {NoStop}%
\bibitem [{\citenamefont {Rozali}\ \emph {et~al.}(2008)\citenamefont {Rozali},
  \citenamefont {Shieh}, \citenamefont {Van~Raamsdonk},\ and\ \citenamefont
  {Wu}}]{Rozali:2007rx}%
  \BibitemOpen
  \bibfield  {author} {\bibinfo {author} {\bibfnamefont {M.}~\bibnamefont
  {Rozali}}, \bibinfo {author} {\bibfnamefont {H.-H.}\ \bibnamefont {Shieh}},
  \bibinfo {author} {\bibfnamefont {M.}~\bibnamefont {Van~Raamsdonk}},\ and\
  \bibinfo {author} {\bibfnamefont {J.}~\bibnamefont {Wu}},\ }\href
  {https://doi.org/10.1088/1126-6708/2008/01/053} {\bibfield  {journal}
  {\bibinfo  {journal} {JHEP}\ }\textbf {\bibinfo {volume} {01}},\ \bibinfo
  {pages} {053}},\ \Eprint {https://arxiv.org/abs/0708.1322} {arXiv:0708.1322
  [hep-th]} \BibitemShut {NoStop}%
\bibitem [{\citenamefont {Kim}\ \emph {et~al.}(2008{\natexlab{b}})\citenamefont
  {Kim}, \citenamefont {Sin},\ and\ \citenamefont {Zahed}}]{Kim:2007vd}%
  \BibitemOpen
  \bibfield  {author} {\bibinfo {author} {\bibfnamefont {K.-Y.}\ \bibnamefont
  {Kim}}, \bibinfo {author} {\bibfnamefont {S.-J.}\ \bibnamefont {Sin}},\ and\
  \bibinfo {author} {\bibfnamefont {I.}~\bibnamefont {Zahed}},\ }\href
  {https://doi.org/10.1088/1126-6708/2008/09/001} {\bibfield  {journal}
  {\bibinfo  {journal} {JHEP}\ }\textbf {\bibinfo {volume} {09}},\ \bibinfo
  {pages} {001}},\ \Eprint {https://arxiv.org/abs/0712.1582} {arXiv:0712.1582
  [hep-th]} \BibitemShut {NoStop}%
\bibitem [{\citenamefont {Sutcliffe}(2010)}]{Sutcliffe:2010et}%
  \BibitemOpen
  \bibfield  {author} {\bibinfo {author} {\bibfnamefont {P.}~\bibnamefont
  {Sutcliffe}},\ }\href {https://doi.org/10.1007/JHEP08(2010)019} {\bibfield
  {journal} {\bibinfo  {journal} {JHEP}\ }\textbf {\bibinfo {volume} {08}},\
  \bibinfo {pages} {019}},\ \Eprint {https://arxiv.org/abs/1003.0023}
  {arXiv:1003.0023 [hep-th]} \BibitemShut {NoStop}%
\bibitem [{\citenamefont {Braga}\ and\ \citenamefont
  {Junqueira}(2024)}]{Braga:2024nnj}%
  \BibitemOpen
  \bibfield  {author} {\bibinfo {author} {\bibfnamefont {N.~R.~F.}\
  \bibnamefont {Braga}}\ and\ \bibinfo {author} {\bibfnamefont {O.~C.}\
  \bibnamefont {Junqueira}},\ }\href
  {https://doi.org/10.1016/j.physletb.2024.138813} {\bibfield  {journal}
  {\bibinfo  {journal} {Phys. Lett. B}\ }\textbf {\bibinfo {volume} {855}},\
  \bibinfo {pages} {138813} (\bibinfo {year} {2024})},\ \Eprint
  {https://arxiv.org/abs/2404.04683} {arXiv:2404.04683 [hep-th]} \BibitemShut
  {NoStop}%
\bibitem [{\citenamefont {Bedaque}\ and\ \citenamefont
  {Steiner}(2015)}]{Bedaque:2014sqa}%
  \BibitemOpen
  \bibfield  {author} {\bibinfo {author} {\bibfnamefont {P.}~\bibnamefont
  {Bedaque}}\ and\ \bibinfo {author} {\bibfnamefont {A.~W.}\ \bibnamefont
  {Steiner}},\ }\href {https://doi.org/10.1103/PhysRevLett.114.031103}
  {\bibfield  {journal} {\bibinfo  {journal} {Phys. Rev. Lett.}\ }\textbf
  {\bibinfo {volume} {114}},\ \bibinfo {pages} {031103} (\bibinfo {year}
  {2015})},\ \Eprint {https://arxiv.org/abs/1408.5116} {arXiv:1408.5116
  [nucl-th]} \BibitemShut {NoStop}%
\bibitem [{\citenamefont {Kojo}\ \emph {et~al.}(2015)\citenamefont {Kojo},
  \citenamefont {Powell}, \citenamefont {Song},\ and\ \citenamefont
  {Baym}}]{Kojo:2014rca}%
  \BibitemOpen
  \bibfield  {author} {\bibinfo {author} {\bibfnamefont {T.}~\bibnamefont
  {Kojo}}, \bibinfo {author} {\bibfnamefont {P.~D.}\ \bibnamefont {Powell}},
  \bibinfo {author} {\bibfnamefont {Y.}~\bibnamefont {Song}},\ and\ \bibinfo
  {author} {\bibfnamefont {G.}~\bibnamefont {Baym}},\ }\href
  {https://doi.org/10.1103/PhysRevD.91.045003} {\bibfield  {journal} {\bibinfo
  {journal} {Phys. Rev. D}\ }\textbf {\bibinfo {volume} {91}},\ \bibinfo
  {pages} {045003} (\bibinfo {year} {2015})},\ \Eprint
  {https://arxiv.org/abs/1412.1108} {arXiv:1412.1108 [hep-ph]} \BibitemShut
  {NoStop}%
\bibitem [{\citenamefont {Tews}\ \emph {et~al.}(2018)\citenamefont {Tews},
  \citenamefont {Carlson}, \citenamefont {Gandolfi},\ and\ \citenamefont
  {Reddy}}]{Tews:2018kmu}%
  \BibitemOpen
  \bibfield  {author} {\bibinfo {author} {\bibfnamefont {I.}~\bibnamefont
  {Tews}}, \bibinfo {author} {\bibfnamefont {J.}~\bibnamefont {Carlson}},
  \bibinfo {author} {\bibfnamefont {S.}~\bibnamefont {Gandolfi}},\ and\
  \bibinfo {author} {\bibfnamefont {S.}~\bibnamefont {Reddy}},\ }\href
  {https://doi.org/10.3847/1538-4357/aac267} {\bibfield  {journal} {\bibinfo
  {journal} {Astrophys. J.}\ }\textbf {\bibinfo {volume} {860}},\ \bibinfo
  {pages} {149} (\bibinfo {year} {2018})},\ \Eprint
  {https://arxiv.org/abs/1801.01923} {arXiv:1801.01923 [nucl-th]} \BibitemShut
  {NoStop}%
\bibitem [{\citenamefont {Ma}\ and\ \citenamefont {Rho}(2019)}]{Ma:2018xjw}%
  \BibitemOpen
  \bibfield  {author} {\bibinfo {author} {\bibfnamefont {Y.-L.}\ \bibnamefont
  {Ma}}\ and\ \bibinfo {author} {\bibfnamefont {M.}~\bibnamefont {Rho}},\
  }\href {https://doi.org/10.1103/PhysRevD.99.014034} {\bibfield  {journal}
  {\bibinfo  {journal} {Phys. Rev. D}\ }\textbf {\bibinfo {volume} {99}},\
  \bibinfo {pages} {014034} (\bibinfo {year} {2019})},\ \Eprint
  {https://arxiv.org/abs/1810.06062} {arXiv:1810.06062 [nucl-th]} \BibitemShut
  {NoStop}%
\bibitem [{\citenamefont {Ma}\ \emph {et~al.}(2019)\citenamefont {Ma},
  \citenamefont {Lee}, \citenamefont {Paeng},\ and\ \citenamefont
  {Rho}}]{Ma:2018jze}%
  \BibitemOpen
  \bibfield  {author} {\bibinfo {author} {\bibfnamefont {Y.-L.}\ \bibnamefont
  {Ma}}, \bibinfo {author} {\bibfnamefont {H.~K.}\ \bibnamefont {Lee}},
  \bibinfo {author} {\bibfnamefont {W.-G.}\ \bibnamefont {Paeng}},\ and\
  \bibinfo {author} {\bibfnamefont {M.}~\bibnamefont {Rho}},\ }\href
  {https://doi.org/10.1007/s11433-019-9399-5} {\bibfield  {journal} {\bibinfo
  {journal} {Sci. China Phys. Mech. Astron.}\ }\textbf {\bibinfo {volume}
  {62}},\ \bibinfo {pages} {112011} (\bibinfo {year} {2019})},\ \Eprint
  {https://arxiv.org/abs/1804.00305} {arXiv:1804.00305 [nucl-th]} \BibitemShut
  {NoStop}%
\bibitem [{\citenamefont {Fujimoto}\ \emph {et~al.}(2022)\citenamefont
  {Fujimoto}, \citenamefont {Fukushima}, \citenamefont {McLerran},\ and\
  \citenamefont {Praszalowicz}}]{Fujimoto:2022ohj}%
  \BibitemOpen
  \bibfield  {author} {\bibinfo {author} {\bibfnamefont {Y.}~\bibnamefont
  {Fujimoto}}, \bibinfo {author} {\bibfnamefont {K.}~\bibnamefont {Fukushima}},
  \bibinfo {author} {\bibfnamefont {L.~D.}\ \bibnamefont {McLerran}},\ and\
  \bibinfo {author} {\bibfnamefont {M.}~\bibnamefont {Praszalowicz}},\ }\href
  {https://doi.org/10.1103/PhysRevLett.129.252702} {\bibfield  {journal}
  {\bibinfo  {journal} {Phys. Rev. Lett.}\ }\textbf {\bibinfo {volume} {129}},\
  \bibinfo {pages} {252702} (\bibinfo {year} {2022})},\ \Eprint
  {https://arxiv.org/abs/2207.06753} {arXiv:2207.06753 [nucl-th]} \BibitemShut
  {NoStop}%
\bibitem [{\citenamefont {Zhang}\ \emph {et~al.}(2024)\citenamefont {Zhang},
  \citenamefont {Ma},\ and\ \citenamefont {Ma}}]{Zhang:2024sju}%
  \BibitemOpen
  \bibfield  {author} {\bibinfo {author} {\bibfnamefont {L.-Q.}\ \bibnamefont
  {Zhang}}, \bibinfo {author} {\bibfnamefont {Y.}~\bibnamefont {Ma}},\ and\
  \bibinfo {author} {\bibfnamefont {Y.-L.}\ \bibnamefont {Ma}},\ }\href@noop {}
  {\  (\bibinfo {year} {2024})},\ \Eprint {https://arxiv.org/abs/2410.04142}
  {arXiv:2410.04142 [nucl-th]} \BibitemShut {NoStop}%
\bibitem [{\citenamefont {Bartz}\ \emph {et~al.}(2024)\citenamefont {Bartz},
  \citenamefont {Meadows},\ and\ \citenamefont {Brock}}]{Bartz:2024dgd}%
  \BibitemOpen
  \bibfield  {author} {\bibinfo {author} {\bibfnamefont {S.~P.}\ \bibnamefont
  {Bartz}}, \bibinfo {author} {\bibfnamefont {R.~C.}\ \bibnamefont {Meadows}},\
  and\ \bibinfo {author} {\bibfnamefont {G.}~\bibnamefont {Brock}},\ }\href
  {https://doi.org/10.1103/PhysRevD.110.026027} {\bibfield  {journal} {\bibinfo
   {journal} {Phys. Rev. D}\ }\textbf {\bibinfo {volume} {110}},\ \bibinfo
  {pages} {026027} (\bibinfo {year} {2024})},\ \Eprint
  {https://arxiv.org/abs/2404.10104} {arXiv:2404.10104 [hep-ph]} \BibitemShut
  {NoStop}%
\bibitem [{\citenamefont {Bartz}\ and\ \citenamefont
  {Jacobson}(2016)}]{Bartz:2016ufc}%
  \BibitemOpen
  \bibfield  {author} {\bibinfo {author} {\bibfnamefont {S.~P.}\ \bibnamefont
  {Bartz}}\ and\ \bibinfo {author} {\bibfnamefont {T.}~\bibnamefont
  {Jacobson}},\ }\href {https://doi.org/10.1103/PhysRevD.94.075022} {\bibfield
  {journal} {\bibinfo  {journal} {Phys. Rev. D}\ }\textbf {\bibinfo {volume}
  {94}},\ \bibinfo {pages} {075022} (\bibinfo {year} {2016})},\ \Eprint
  {https://arxiv.org/abs/1607.05751} {arXiv:1607.05751 [hep-ph]} \BibitemShut
  {NoStop}%
\bibitem [{Note1()}]{Note1}%
  \BibitemOpen
  \bibinfo {note} {At zero density, chiral condensation $\Sigma $ and $\omega
  _{ir}$ are equivalent. At finite density, although they are not strictly
  equivalent, $\omega _{ir}$ can still be considered as an order parameter for
  whether the chiral condensation is zero.}\BibitemShut {Stop}%
\bibitem [{\citenamefont {Bartolini}\ and\ \citenamefont
  {Gudnason}(2023)}]{Bartolini:2023wis}%
  \BibitemOpen
  \bibfield  {author} {\bibinfo {author} {\bibfnamefont {L.}~\bibnamefont
  {Bartolini}}\ and\ \bibinfo {author} {\bibfnamefont {S.~B.}\ \bibnamefont
  {Gudnason}},\ }\href {https://doi.org/10.1007/JHEP11(2023)209} {\bibfield
  {journal} {\bibinfo  {journal} {Journal of High Energy Physics}\ }\textbf
  {\bibinfo {volume} {2023}},\ \bibinfo {pages} {209} (\bibinfo {year}
  {2023})},\ \Eprint {https://arxiv.org/abs/2307.11886} {arXiv:2307.11886
  [hep-ph, physics:nucl-th]} \BibitemShut {NoStop}%
\bibitem [{\citenamefont {Witten}(1979)}]{Witten:1979kh}%
  \BibitemOpen
  \bibfield  {author} {\bibinfo {author} {\bibfnamefont {E.}~\bibnamefont
  {Witten}},\ }\href {https://doi.org/10.1016/0550-3213(79)90232-3} {\bibfield
  {journal} {\bibinfo  {journal} {Nucl. Phys. B}\ }\textbf {\bibinfo {volume}
  {160}},\ \bibinfo {pages} {57} (\bibinfo {year} {1979})}\BibitemShut
  {NoStop}%
\bibitem [{\citenamefont {Ma}\ \emph {et~al.}(2013)\citenamefont {Ma},
  \citenamefont {Harada}, \citenamefont {Lee}, \citenamefont {Oh},
  \citenamefont {Park},\ and\ \citenamefont {Rho}}]{Ma:2013ooa}%
  \BibitemOpen
  \bibfield  {author} {\bibinfo {author} {\bibfnamefont {Y.-L.}\ \bibnamefont
  {Ma}}, \bibinfo {author} {\bibfnamefont {M.}~\bibnamefont {Harada}}, \bibinfo
  {author} {\bibfnamefont {H.~K.}\ \bibnamefont {Lee}}, \bibinfo {author}
  {\bibfnamefont {Y.}~\bibnamefont {Oh}}, \bibinfo {author} {\bibfnamefont
  {B.-Y.}\ \bibnamefont {Park}},\ and\ \bibinfo {author} {\bibfnamefont
  {M.}~\bibnamefont {Rho}},\ }\href
  {https://doi.org/10.1103/PhysRevD.88.014016} {\bibfield  {journal} {\bibinfo
  {journal} {Phys. Rev. D}\ }\textbf {\bibinfo {volume} {88}},\ \bibinfo
  {pages} {014016} (\bibinfo {year} {2013})},\ \bibinfo {note} {[Erratum:
  Phys.Rev.D 88, 079904 (2013)]},\ \Eprint {https://arxiv.org/abs/1304.5638}
  {arXiv:1304.5638 [hep-ph]} \BibitemShut {NoStop}%
\bibitem [{\citenamefont {Ma}\ \emph {et~al.}(2014)\citenamefont {Ma},
  \citenamefont {Harada}, \citenamefont {Lee}, \citenamefont {Oh},
  \citenamefont {Park},\ and\ \citenamefont {Rho}}]{Ma:2013ela}%
  \BibitemOpen
  \bibfield  {author} {\bibinfo {author} {\bibfnamefont {Y.-L.}\ \bibnamefont
  {Ma}}, \bibinfo {author} {\bibfnamefont {M.}~\bibnamefont {Harada}}, \bibinfo
  {author} {\bibfnamefont {H.~K.}\ \bibnamefont {Lee}}, \bibinfo {author}
  {\bibfnamefont {Y.}~\bibnamefont {Oh}}, \bibinfo {author} {\bibfnamefont
  {B.-Y.}\ \bibnamefont {Park}},\ and\ \bibinfo {author} {\bibfnamefont
  {M.}~\bibnamefont {Rho}},\ }\href
  {https://doi.org/10.1103/PhysRevD.90.034015} {\bibfield  {journal} {\bibinfo
  {journal} {Phys. Rev. D}\ }\textbf {\bibinfo {volume} {90}},\ \bibinfo
  {pages} {034015} (\bibinfo {year} {2014})},\ \Eprint
  {https://arxiv.org/abs/1308.6476} {arXiv:1308.6476 [hep-ph]} \BibitemShut
  {NoStop}%
\bibitem [{\citenamefont {Shao}\ and\ \citenamefont {Ma}(2022)}]{Shao:2022njr}%
  \BibitemOpen
  \bibfield  {author} {\bibinfo {author} {\bibfnamefont {L.-Q.}\ \bibnamefont
  {Shao}}\ and\ \bibinfo {author} {\bibfnamefont {Y.-L.}\ \bibnamefont {Ma}},\
  }\href {https://doi.org/10.1103/PhysRevD.106.014014} {\bibfield  {journal}
  {\bibinfo  {journal} {Phys. Rev. D}\ }\textbf {\bibinfo {volume} {106}},\
  \bibinfo {pages} {014014} (\bibinfo {year} {2022})},\ \Eprint
  {https://arxiv.org/abs/2202.09957} {arXiv:2202.09957 [nucl-th]} \BibitemShut
  {NoStop}%
\bibitem [{\citenamefont {Chabanat}\ \emph {et~al.}(1997)\citenamefont
  {Chabanat}, \citenamefont {Meyer}, \citenamefont {Bonche}, \citenamefont
  {Schaeffer},\ and\ \citenamefont {Haensel}}]{Chabanat:1997qh}%
  \BibitemOpen
  \bibfield  {author} {\bibinfo {author} {\bibfnamefont {E.}~\bibnamefont
  {Chabanat}}, \bibinfo {author} {\bibfnamefont {J.}~\bibnamefont {Meyer}},
  \bibinfo {author} {\bibfnamefont {P.}~\bibnamefont {Bonche}}, \bibinfo
  {author} {\bibfnamefont {R.}~\bibnamefont {Schaeffer}},\ and\ \bibinfo
  {author} {\bibfnamefont {P.}~\bibnamefont {Haensel}},\ }\href
  {https://doi.org/10.1016/S0375-9474(97)00596-4} {\bibfield  {journal}
  {\bibinfo  {journal} {Nucl. Phys. A}\ }\textbf {\bibinfo {volume} {627}},\
  \bibinfo {pages} {710} (\bibinfo {year} {1997})}\BibitemShut {NoStop}%
\bibitem [{\citenamefont {Chabanat}\ \emph {et~al.}(1998)\citenamefont
  {Chabanat}, \citenamefont {Bonche}, \citenamefont {Haensel}, \citenamefont
  {Meyer},\ and\ \citenamefont {Schaeffer}}]{Chabanat:1997un}%
  \BibitemOpen
  \bibfield  {author} {\bibinfo {author} {\bibfnamefont {E.}~\bibnamefont
  {Chabanat}}, \bibinfo {author} {\bibfnamefont {P.}~\bibnamefont {Bonche}},
  \bibinfo {author} {\bibfnamefont {P.}~\bibnamefont {Haensel}}, \bibinfo
  {author} {\bibfnamefont {J.}~\bibnamefont {Meyer}},\ and\ \bibinfo {author}
  {\bibfnamefont {R.}~\bibnamefont {Schaeffer}},\ }\href
  {https://doi.org/10.1016/S0375-9474(98)00180-8} {\bibfield  {journal}
  {\bibinfo  {journal} {Nucl. Phys. A}\ }\textbf {\bibinfo {volume} {635}},\
  \bibinfo {pages} {231} (\bibinfo {year} {1998})},\ \bibinfo {note} {[Erratum:
  Nucl.Phys.A 643, 441--441 (1998)]}\BibitemShut {NoStop}%
\bibitem [{\citenamefont {Douchin}\ and\ \citenamefont
  {Haensel}(2000)}]{Douchin:2000kx}%
  \BibitemOpen
  \bibfield  {author} {\bibinfo {author} {\bibfnamefont {F.}~\bibnamefont
  {Douchin}}\ and\ \bibinfo {author} {\bibfnamefont {P.}~\bibnamefont
  {Haensel}},\ }\href {https://doi.org/10.1016/S0370-2693(00)00672-9}
  {\bibfield  {journal} {\bibinfo  {journal} {Phys. Lett. B}\ }\textbf
  {\bibinfo {volume} {485}},\ \bibinfo {pages} {107} (\bibinfo {year}
  {2000})},\ \Eprint {https://arxiv.org/abs/astro-ph/0006135}
  {arXiv:astro-ph/0006135} \BibitemShut {NoStop}%
\bibitem [{\citenamefont {Baym}\ \emph {et~al.}(1971)\citenamefont {Baym},
  \citenamefont {Pethick},\ and\ \citenamefont {Sutherland}}]{Baym:1971pw}%
  \BibitemOpen
  \bibfield  {author} {\bibinfo {author} {\bibfnamefont {G.}~\bibnamefont
  {Baym}}, \bibinfo {author} {\bibfnamefont {C.}~\bibnamefont {Pethick}},\ and\
  \bibinfo {author} {\bibfnamefont {P.}~\bibnamefont {Sutherland}},\ }\href
  {https://doi.org/10.1086/151216} {\bibfield  {journal} {\bibinfo  {journal}
  {Astrophys. J.}\ }\textbf {\bibinfo {volume} {170}},\ \bibinfo {pages} {299}
  (\bibinfo {year} {1971})}\BibitemShut {NoStop}%
\bibitem [{Note2()}]{Note2}%
  \BibitemOpen
  \bibinfo {note} {Https://github.com/amotornenko/TOVsolver,
  https://github.com/jnoronhahostler/{Neutron\protect \_Star\protect
  \_EOS}}\BibitemShut {NoStop}%
\bibitem [{\citenamefont {Tan}\ \emph {et~al.}(2020)\citenamefont {Tan},
  \citenamefont {Noronha-Hostler},\ and\ \citenamefont {Yunes}}]{Tan:2020ics}%
  \BibitemOpen
  \bibfield  {author} {\bibinfo {author} {\bibfnamefont {H.}~\bibnamefont
  {Tan}}, \bibinfo {author} {\bibfnamefont {J.}~\bibnamefont
  {Noronha-Hostler}},\ and\ \bibinfo {author} {\bibfnamefont {N.}~\bibnamefont
  {Yunes}},\ }\href {https://doi.org/10.1103/PhysRevLett.125.261104} {\bibfield
   {journal} {\bibinfo  {journal} {Phys. Rev. Lett.}\ }\textbf {\bibinfo
  {volume} {125}},\ \bibinfo {pages} {261104} (\bibinfo {year} {2020})},\
  \Eprint {https://arxiv.org/abs/2006.16296} {arXiv:2006.16296 [astro-ph.HE]}
  \BibitemShut {NoStop}%
\bibitem [{\citenamefont {Miller}\ \emph {et~al.}(2019)\citenamefont {Miller}
  \emph {et~al.}}]{Miller:2019cac}%
  \BibitemOpen
  \bibfield  {author} {\bibinfo {author} {\bibfnamefont {M.~C.}\ \bibnamefont
  {Miller}} \emph {et~al.},\ }\href {https://doi.org/10.3847/2041-8213/ab50c5}
  {\bibfield  {journal} {\bibinfo  {journal} {Astrophys. J. Lett.}\ }\textbf
  {\bibinfo {volume} {887}},\ \bibinfo {pages} {L24} (\bibinfo {year}
  {2019})},\ \Eprint {https://arxiv.org/abs/1912.05705} {arXiv:1912.05705
  [astro-ph.HE]} \BibitemShut {NoStop}%
\bibitem [{\citenamefont {Abbott}\ \emph {et~al.}(2020)\citenamefont {Abbott}
  \emph {et~al.}}]{LIGOScientific:2020zkf}%
  \BibitemOpen
  \bibfield  {author} {\bibinfo {author} {\bibfnamefont {R.}~\bibnamefont
  {Abbott}} \emph {et~al.} (\bibinfo {collaboration} {LIGO Scientific,
  Virgo}),\ }\href {https://doi.org/10.3847/2041-8213/ab960f} {\bibfield
  {journal} {\bibinfo  {journal} {Astrophys. J. Lett.}\ }\textbf {\bibinfo
  {volume} {896}},\ \bibinfo {pages} {L44} (\bibinfo {year} {2020})},\ \Eprint
  {https://arxiv.org/abs/2006.12611} {arXiv:2006.12611 [astro-ph.HE]}
  \BibitemShut {NoStop}%
\bibitem [{\citenamefont {Abbott}\ \emph {et~al.}(2018)\citenamefont {Abbott}
  \emph {et~al.}}]{LIGOScientific:2018cki}%
  \BibitemOpen
  \bibfield  {author} {\bibinfo {author} {\bibfnamefont {B.~P.}\ \bibnamefont
  {Abbott}} \emph {et~al.} (\bibinfo {collaboration} {LIGO Scientific,
  Virgo}),\ }\href {https://doi.org/10.1103/PhysRevLett.121.161101} {\bibfield
  {journal} {\bibinfo  {journal} {Phys. Rev. Lett.}\ }\textbf {\bibinfo
  {volume} {121}},\ \bibinfo {pages} {161101} (\bibinfo {year} {2018})},\
  \Eprint {https://arxiv.org/abs/1805.11581} {arXiv:1805.11581 [gr-qc]}
  \BibitemShut {NoStop}%
\bibitem [{\citenamefont {Miller}\ \emph {et~al.}(2021)\citenamefont {Miller}
  \emph {et~al.}}]{Miller:2021qha}%
  \BibitemOpen
  \bibfield  {author} {\bibinfo {author} {\bibfnamefont {M.~C.}\ \bibnamefont
  {Miller}} \emph {et~al.},\ }\href {https://doi.org/10.3847/2041-8213/ac089b}
  {\bibfield  {journal} {\bibinfo  {journal} {Astrophys. J. Lett.}\ }\textbf
  {\bibinfo {volume} {918}},\ \bibinfo {pages} {L28} (\bibinfo {year}
  {2021})},\ \Eprint {https://arxiv.org/abs/2105.06979} {arXiv:2105.06979
  [astro-ph.HE]} \BibitemShut {NoStop}%
\bibitem [{\citenamefont {Riley}\ \emph {et~al.}(2021)\citenamefont {Riley}
  \emph {et~al.}}]{Riley:2021pdl}%
  \BibitemOpen
  \bibfield  {author} {\bibinfo {author} {\bibfnamefont {T.~E.}\ \bibnamefont
  {Riley}} \emph {et~al.},\ }\href {https://doi.org/10.3847/2041-8213/ac0a81}
  {\bibfield  {journal} {\bibinfo  {journal} {Astrophys. J. Lett.}\ }\textbf
  {\bibinfo {volume} {918}},\ \bibinfo {pages} {L27} (\bibinfo {year}
  {2021})},\ \Eprint {https://arxiv.org/abs/2105.06980} {arXiv:2105.06980
  [astro-ph.HE]} \BibitemShut {NoStop}%
\bibitem [{\citenamefont {Sedrakian}\ \emph {et~al.}(2023)\citenamefont
  {Sedrakian}, \citenamefont {Li},\ and\ \citenamefont
  {Weber}}]{Sedrakian:2022ata}%
  \BibitemOpen
  \bibfield  {author} {\bibinfo {author} {\bibfnamefont {A.}~\bibnamefont
  {Sedrakian}}, \bibinfo {author} {\bibfnamefont {J.-J.}\ \bibnamefont {Li}},\
  and\ \bibinfo {author} {\bibfnamefont {F.}~\bibnamefont {Weber}},\ }\href
  {https://doi.org/10.1016/j.ppnp.2023.104041} {\bibfield  {journal} {\bibinfo
  {journal} {Prog. Part. Nucl. Phys.}\ }\textbf {\bibinfo {volume} {131}},\
  \bibinfo {pages} {104041} (\bibinfo {year} {2023})},\ \Eprint
  {https://arxiv.org/abs/2212.01086} {arXiv:2212.01086 [nucl-th]} \BibitemShut
  {NoStop}%
\bibitem [{\citenamefont {Kovensky}\ \emph {et~al.}(2023)\citenamefont
  {Kovensky}, \citenamefont {Poole},\ and\ \citenamefont
  {Schmitt}}]{Kovensky:2023mye}%
  \BibitemOpen
  \bibfield  {author} {\bibinfo {author} {\bibfnamefont {N.}~\bibnamefont
  {Kovensky}}, \bibinfo {author} {\bibfnamefont {A.}~\bibnamefont {Poole}},\
  and\ \bibinfo {author} {\bibfnamefont {A.}~\bibnamefont {Schmitt}},\ }\href
  {https://doi.org/10.21468/SciPostPhys.15.4.162} {\bibfield  {journal}
  {\bibinfo  {journal} {SciPost Phys.}\ }\textbf {\bibinfo {volume} {15}},\
  \bibinfo {pages} {162} (\bibinfo {year} {2023})},\ \Eprint
  {https://arxiv.org/abs/2302.10675} {arXiv:2302.10675 [hep-ph]} \BibitemShut
  {NoStop}%
\bibitem [{\citenamefont {Bartolini}\ \emph {et~al.}(2025)\citenamefont
  {Bartolini}, \citenamefont {Gudnason},\ and\ \citenamefont
  {J{\"a}rvinen}}]{Bartolini:2025sag}%
  \BibitemOpen
  \bibfield  {author} {\bibinfo {author} {\bibfnamefont {L.}~\bibnamefont
  {Bartolini}}, \bibinfo {author} {\bibfnamefont {S.~B.}\ \bibnamefont
  {Gudnason}},\ and\ \bibinfo {author} {\bibfnamefont {M.}~\bibnamefont
  {J{\"a}rvinen}},\ }\href {https://doi.org/10.1103/PhysRevD.111.106021}
  {\bibfield  {journal} {\bibinfo  {journal} {Phys. Rev. D}\ }\textbf {\bibinfo
  {volume} {111}},\ \bibinfo {pages} {106021} (\bibinfo {year} {2025})},\
  \Eprint {https://arxiv.org/abs/2504.01758} {arXiv:2504.01758 [hep-ph]}
  \BibitemShut {NoStop}%
\bibitem [{\citenamefont {Weigel}(2008)}]{weigelChiralSolitonModels2008}%
  \BibitemOpen
  \bibfield  {author} {\bibinfo {author} {\bibfnamefont {H.}~\bibnamefont
  {Weigel}},\ }\href@noop {} {\emph {\bibinfo {title} {Chiral Soliton Models
  for Baryons}}},\ \bibinfo {series} {Lecture Notes in Physics}\ No.\ \bibinfo
  {number} {743}\ (\bibinfo  {publisher} {Springer},\ \bibinfo {address}
  {Berlin New York},\ \bibinfo {year} {2008})\BibitemShut {NoStop}%
\bibitem [{\citenamefont {Manton}(2022)}]{mantonSkyrmionsTheoryNuclei2022}%
  \BibitemOpen
  \bibfield  {author} {\bibinfo {author} {\bibfnamefont {N.}~\bibnamefont
  {Manton}},\ }\href@noop {} {\emph {\bibinfo {title} {Skyrmions: A Theory of
  Nuclei}}}\ (\bibinfo  {publisher} {World scientific},\ \bibinfo {address}
  {London},\ \bibinfo {year} {2022})\BibitemShut {NoStop}%
\bibitem [{\citenamefont {Harada}\ \emph {et~al.}(2010)\citenamefont {Harada},
  \citenamefont {Matsuzaki},\ and\ \citenamefont
  {Yamawaki}}]{PhysRevD.82.076010}%
  \BibitemOpen
  \bibfield  {author} {\bibinfo {author} {\bibfnamefont {M.}~\bibnamefont
  {Harada}}, \bibinfo {author} {\bibfnamefont {S.}~\bibnamefont {Matsuzaki}},\
  and\ \bibinfo {author} {\bibfnamefont {K.}~\bibnamefont {Yamawaki}},\ }\href
  {https://doi.org/10.1103/PhysRevD.82.076010} {\bibfield  {journal} {\bibinfo
  {journal} {Phys. Rev. D}\ }\textbf {\bibinfo {volume} {82}},\ \bibinfo
  {pages} {076010} (\bibinfo {year} {2010})}\BibitemShut {NoStop}%
\bibitem [{\citenamefont {Harada}\ \emph {et~al.}(2014)\citenamefont {Harada},
  \citenamefont {Ma},\ and\ \citenamefont {Matsuzaki}}]{PhysRevD.89.115012}%
  \BibitemOpen
  \bibfield  {author} {\bibinfo {author} {\bibfnamefont {M.}~\bibnamefont
  {Harada}}, \bibinfo {author} {\bibfnamefont {Y.-L.}\ \bibnamefont {Ma}},\
  and\ \bibinfo {author} {\bibfnamefont {S.}~\bibnamefont {Matsuzaki}},\ }\href
  {https://doi.org/10.1103/PhysRevD.89.115012} {\bibfield  {journal} {\bibinfo
  {journal} {Phys. Rev. D}\ }\textbf {\bibinfo {volume} {89}},\ \bibinfo
  {pages} {115012} (\bibinfo {year} {2014})}\BibitemShut {NoStop}%
\bibitem [{\citenamefont {Harada}\ and\ \citenamefont
  {Yamawaki}(2003)}]{Harada:2003jx}%
  \BibitemOpen
  \bibfield  {author} {\bibinfo {author} {\bibfnamefont {M.}~\bibnamefont
  {Harada}}\ and\ \bibinfo {author} {\bibfnamefont {K.}~\bibnamefont
  {Yamawaki}},\ }\href {https://doi.org/10.1016/S0370-1573(03)00139-X}
  {\bibfield  {journal} {\bibinfo  {journal} {Phys. Rept.}\ }\textbf {\bibinfo
  {volume} {381}},\ \bibinfo {pages} {1} (\bibinfo {year} {2003})},\ \Eprint
  {https://arxiv.org/abs/hep-ph/0302103} {arXiv:hep-ph/0302103} \BibitemShut
  {NoStop}%
\bibitem [{\citenamefont {Pomarol}\ and\ \citenamefont
  {Wulzer}(2009)}]{Pomarol:2008aa}%
  \BibitemOpen
  \bibfield  {author} {\bibinfo {author} {\bibfnamefont {A.}~\bibnamefont
  {Pomarol}}\ and\ \bibinfo {author} {\bibfnamefont {A.}~\bibnamefont
  {Wulzer}},\ }\href {https://doi.org/10.1016/j.nuclphysb.2008.10.004}
  {\bibfield  {journal} {\bibinfo  {journal} {Nucl. Phys. B}\ }\textbf
  {\bibinfo {volume} {809}},\ \bibinfo {pages} {347} (\bibinfo {year}
  {2009})},\ \Eprint {https://arxiv.org/abs/0807.0316} {arXiv:0807.0316
  [hep-ph]} \BibitemShut {NoStop}%
\bibitem [{\citenamefont {Bhattacharyya}\ \emph {et~al.}(2010)\citenamefont
  {Bhattacharyya}, \citenamefont {Mishustin},\ and\ \citenamefont
  {Greiner}}]{Bhattacharyya:2009fg}%
  \BibitemOpen
  \bibfield  {author} {\bibinfo {author} {\bibfnamefont {A.}~\bibnamefont
  {Bhattacharyya}}, \bibinfo {author} {\bibfnamefont {I.~N.}\ \bibnamefont
  {Mishustin}},\ and\ \bibinfo {author} {\bibfnamefont {W.}~\bibnamefont
  {Greiner}},\ }\href {https://doi.org/10.1088/0954-3899/37/2/025201}
  {\bibfield  {journal} {\bibinfo  {journal} {J. Phys. G}\ }\textbf {\bibinfo
  {volume} {37}},\ \bibinfo {pages} {025201} (\bibinfo {year} {2010})},\
  \Eprint {https://arxiv.org/abs/0905.0352} {arXiv:0905.0352 [nucl-th]}
  \BibitemShut {NoStop}%
\bibitem [{\citenamefont {Masuda}\ \emph {et~al.}(2013)\citenamefont {Masuda},
  \citenamefont {Hatsuda},\ and\ \citenamefont {Takatsuka}}]{Masuda:2012ed}%
  \BibitemOpen
  \bibfield  {author} {\bibinfo {author} {\bibfnamefont {K.}~\bibnamefont
  {Masuda}}, \bibinfo {author} {\bibfnamefont {T.}~\bibnamefont {Hatsuda}},\
  and\ \bibinfo {author} {\bibfnamefont {T.}~\bibnamefont {Takatsuka}},\ }\href
  {https://doi.org/10.1093/ptep/ptt045} {\bibfield  {journal} {\bibinfo
  {journal} {PTEP}\ }\textbf {\bibinfo {volume} {2013}},\ \bibinfo {pages}
  {073D01} (\bibinfo {year} {2013})},\ \Eprint
  {https://arxiv.org/abs/1212.6803} {arXiv:1212.6803 [nucl-th]} \BibitemShut
  {NoStop}%
\end{thebibliography}%
\end{document}